# A comprehensive review on topological superconducting materials and interfaces

M. M. Sharma[1,2], Prince Sharma[1, 2], N.K. Karn[1,2], V.P.S. Awana[1, 2*]

[1]CSIR-National Physical Laboratory, K.S. Krishnan Marg, New Delhi-110012, India
[2]Academy council of scientific and industrial Research, Ghaziabad U.P.-201002, India

**Abstract:**

Superconductivity in topological materials has drawn a significant interest of the scientific community as these materials provide a hint of existence of Majorana fermions conceived from the quantized thermal conductivity, a zero-biased conduction peak and the anomalous Josephson effect. In this review, we make a systematic study of recent advances in the field of topological superconductivity. The article comprises of both bulk systems as well as heterostructures. A brief description of Majorana fermions and their relationship with topological superconductors and heterostructures is also carried out. Also, this review consists of details of key experimental techniques to characterize candidates of topological superconductivity. Moreover, we summarize the potential material candidate that may demonstrate topological superconductivity. We also consider some intrinsic odd-parity superconductors, non-centrosymmetric, centrosymmetric superconductors, doped topological insulators, doped topological crystalline insulators and some other materials that are expected to show superconductivity along with topological non-trivial states in bulk from. The effect of pressure, emergence of superconductivity in topological materials and Muon Spin Rotation (µSR) studies are also summarized in this article.

Key Words: Superconductivity, Topological Insulators, Majorana fermions, Topological superconductors, and Topological crystalline insulators.

PACS: 74.70.Dd, 74.62.Fj, 74.25.F-

*Corresponding Author

Dr. V. P. S. Awana:  E-mail: awana@nplindia.org
Ph. +91-11-45609357, Fax-+91-11-45609310
*Homepage: awanavps.webs.com*

**Introduction**

A typical class of crystalline solids, magnets and superconductors that have some special states come under the class of quantum matter. These states can be classified by studying their spontaneous symmetry breaking [1]. There exists an order parameter that has a non-zero expectation value in an ordered state. This order parameter can be determined by Ginzburg Landau (G-L) theory and gives a unique description of these quantum state [2-5]. However, in topology, a quantum matter with an order parameter creates a new quantum state, which is beyond the paradigm of Ginzburg Landau Theory. In the 1980s, Von Klitzing discovered such kind of quantum states named Quantum Hall (QH) states in the two-dimensional system. The core of this system has an insulating gap while the edge is conducting due to the skipping orbits that are formed due to the externally applied magnetic field [6]. These edge states provide a unidirectional dissipation less path to electric current, which gives rise to Quantized Hall Effect (QHE) [6]. The topological phases have certain parameters that are insensitive to the smooth deformation in the Hamiltonian, named as Topological invariants. In the surface states, smooth deformation can be classified as the change which does not disturb the bulk energy gap. Moreover, the energy gap in bulk also plays a significant role in the particular topological phases, which gives rise to two new classes of quantum materials named Topological Insulators (TIs) and Topological Superconductor (TSC). In particular, this effect can't be observed in conductors and doped semiconductors as they do not have any energy gap [7].

Topological Insulators were first theoretically predicted in 2005 by C.L. Kane et.al. [8] and these were experimentally realized in 2007 by Konig et.al. [9] in HgTe/CdTe quantum wells. Topological Insulators are different from Quantum Hall systems in a way that the time-reversal symmetry (TRS) is broken in presence of impurity in the latter one. While in case of TIs, TRS is preserved as the spin-orbit interaction plays a vital role. TIs are also termed as Quantum Spin Hall (QSH) systems due to presence of intrinsic spin-orbit coupling (SOC). A QSH state is topologically nontrivial and consists of a fully insulating bulk and conducting surface states with an odd number of Dirac fermions [10-12]. The states containing odd number of fermions always have at least double degeneracy [13], which means that the surface states are paired in topological non-trivial system. These two states propagate in the opposite direction to each other [14].

Topological superconductors are supposed to have greater importance in fault-tolerant quantum computing because of the emergence of the Majorana fermions [15]. TSC can be classified according to their type of superconductivity and the presence of time-reversal symmetry. In the prior one, TSC was categorized as strong and weak TSC. Some unconventional superconductors have a nodal superconducting gap. If these nodal superconductors have topological properties, then these can be regarded as weak TSC [16]. The strong TSC shows three main characteristics that are (1) a full superconducting gap with an odd parity pairing symmetry (2) conducting gapless surface states and (3) Majorana Zero Modes (MZMs) in superconducting vortex cores [16]. In the context of time-reversal symmetry, TSC can also be classified into two categories one in which TRS is broken and the other in which TRS is preserved [17,18]. Generally, superconductors with odd parity pairing have an internal magnetic field which is responsible for the breaking of TRS in the former type of TSC, one such example is $Sr_2RuO_4$ [19-21]. While in the subsequent class, the TRS is preserved due to presence of strong SOC, one such example is $Cu_xBi_2Se_3$ [22].

The surface states of TSC are supposed to contain Majorana fermions [23]. Majorana fermions are different from Dirac fermions. Dirac fermions are the particles that have a separate antiparticle while Majorana fermions are the antiparticles of itself [24]. Realization of Majorana fermions in materials remained an unanswered question until the discovery of topological superconductors [23]. There are several methods to obtain topological superconductivity such as doping in TIs [22], applying pressure in TIs [25] and employing the proximity effect [4,26,27]. In proximity induced topological superconductivity, TI is placed in proximity of an s wave superconductor and, the Dirac fermions present on the surface of TI show superconductivity [26]. Another method to induce topological superconductivity in TIs is by doping of certain elements [22]. The first superconducting doped TI came into the picture, when $Bi_2Se_3$ was found to show superconductivity due to intercalation of Cu between its two quintuple layers [22]. Basically, $Bi_2Se_3$ does not have sufficient charge carriers to show superconductivity, however intercalation of material such as Cu, increases the charge carriers in the same material. This discovery opened the door for the search for doped topological superconductors. Also, application of pressure found to induce superconductivity in some TIs such as $Bi_2Te_3$ shows superconductivity near 2.7K at a pressure of 7GPa [25].

In this review article, we'll briefly discuss Majorana fermions and why they are predicted to be realized in topological superconductors. The characteristics of topological superconductors which indicates the presence of Majorana fermions are also deliberated. Some important experimental techniques to characterize the samples in context of their topological and superconducting properties are also briefly discussed. Among these, the μSR studies are the most suitable technique to study superconducting states and symmetries, and are separately discussed just after the description of bulk topological superconductors. The main focus of this review is the materials that are possible candidates of topological superconductivity. In this regard, we discuss bulk materials that show topological superconductivity in an intrinsic form, along with the bulk TIs which can be made superconducting by doping or by application of pressure. The last section of this review is dedicated to some heterostructures which are supposed to have Majorana fermions at their interface.

**Majorana Fermions & Topological Superconductivity:**

Topological superconductors are considered to be the best possible platform to realize Majorana fermions [15,16,23]. Although not yet detected, TSCs are still very fascinating for both experimental and fundamental research. Here we'll briefly discuss Majorana fermions and the novel characteristics of TSC by which the contemplation of the existence of Majorana fermions had predicted. In 1929, Paul Dirac derived the equation for the relativistic motion of electron known as the Dirac equation. This theory predicted that for each fermion, there exists an antiparticle that is different from particle only in terms of the sign of charge. This prediction was experimentally confirmed in 1934, when Anderson searched a new particle positron that has the same mass and spin as an electron but has a positive charge. In 1937, Attore Majorana predicted the existence of such particles which are identical to their antiparticle in all aspects [24], this prediction was made by using the Dirac equation too. In his work, he derived that complex conjugate of the wave function that satisfies the Dirac equation also satisfies the same without any incongruity. The complex conjugate of the wavefunction that satisfies the Dirac equation is considered to be the wave function of antiparticle. If this wave function is real, then it will be self-conjugate and makes ground for the existence of particles that are similar to their antiparticles. This particular class of particles which are their own antiparticles are known as Majorana particles. These particles follow Fermi Dirac statistics, that's why also knowns as Majorana fermions.

After the theoretical prediction of Majorana fermions, the debate started on the systems in which such particles can be realized. In particle physics, only neutrinos are the particles that fulfills the conditions to be Majorana fermions, but still it is a contentious point whether these can be treated as Majorana fermions or not. There are two necessary conditions for a fermion to be treated as Majorana fermion, one is it should satisfy Dirac relativistic equation and the other is, it should be its own antiparticle. An electron that is an integral fermion in condensed matter, does not satisfy the above conditions, as it obeys the non-relativistic Schrodinger equation and also it has a negative charge which nullifies all possibility of an electron to be its own antiparticle. Thereby, apart from neutrinos, the conditions that a particle must satisfy to be a Majorana fermion can be fulfilled in superconducting topological materials i.e., in TSCs [28]. The possibility of realization of Majorana fermions is based on the two facts: (i) TSCs have topological non-trivial surface states that host gapless excitations and, are governed by the relativistic Dirac equation. (ii) The second is the presence of superconductivity. In TSCs, the superimposition of electron and hole excitations takes place which makes them indistinguishable and charge-neutral enabling them to be their antiparticle [28] as shown in Fig. 1. So TSCs are thought to be important materials to experimentally realize these Majorana fermions. Unconventional odd-parity superconductors (p wave superconductors) are also important in this context, as cooper pairs in p wave superconductors have orbital angular momentum to be 1 and it is highly assumed that Majorana fermions can be observed in the vortex core of these systems too [28,29].

**Signature of Majorana Fermions in Topological Superconductors:**

Existence of Majorana fermions in a TSC can be checked by observing some signatures of the same. In this section we'll have a brief discussion about these signatures:

- **Zero Bias Conduction Peak (ZBCP):**

Topological superconductors are supposed to show bulk superconductivity with topological non-trivial surface states or gapless surface states. These gapless surface states can be probed by point contact spectroscopy, as the same have proved to be a great tool to study the surface density of states. Also, point contact spectroscopy measurements are considered to be a preliminary check of whether the conducting surface states of a topological superconductor hosts Majorana fermions or not. The presence of Majorana fermions in conducting surface states imposes a significant impact on conductance between the normal metallic and superconducting states. This leads to a peak in conduction spectra at zero bias due to tunneling of the electron as suggested in various reports [30-32]. While in the case of unconventional superconductors, Majorana fermions are expected to be present in their vortex core [29]. This would be demonstrated by a zero biased peak in the conductance spectra of the point contact spectroscopy measurements, that are made inside the vortex core [33]. But the presence of zero biased peaks is only a preliminary check. It alone does not confirm the presence of Majorana fermions in the system as there can be other reasons by which this peak may be seen such as demonstrating heating effect at the contact [34], reflection less tunneling [35,36] or magnetic scattering [37]. In particular a continuum of Caroli-Matricon-deGenne states leads to a zero-bias peak in the vortex core [38]. One has to check whether ZBCP is occurred due to these effects or due to the presence of Majorana modes, but all in all, it is one of a preliminary check of the presence of Majorana fermions in surface states.

- **Quantized Thermal Conductivity:**

The presence of Majorana fermions in the surface states leads to an exotic phenomenon known as Thermal Quantum Hall Effect [17]. In a topological superconductor, the bulk is superconducting which remains thermally insulating due to which, Majorana bound states appear too distant to carry any heat. Moreover, only the surface states are available for heat conduction [17,39]. Fig. 2 shows a pictorial view of thermal hall conductance due to Majorana edge states in a p wave superconductor. Similar to the Quantum Hall Effect, there is a quantization unit of thermal Quantum Hall Effect and it is given by $\frac{1}{3}(\pi k_b/e)^2 T e^2/h$, where T is temperature and $\frac{1}{3}(\pi k_b/e)^2$ is Lorentz number [39].

- **Anomalous Josephson Effect:**

Another interesting feature that is considered to be a consequence of the presence of Majorana fermions is the anomalous Josephson Effect. It differs from ZBCP because in the case of ZBCP a single electron tunnels from the superconducting state to Majorana states, while an anomalous Josephson Effect occurs due to the tunneling of Cooper pairs [41]. This effect occurs between a junction of two topological superconductors. It is observed that when a Cooper pair tunnels from one TSc to another going through Majorana bound states with the phase relationship between voltage and current changes from 2π to 4π, leading to a fractional Josephson effect. This fractional Josephson effect is known as anomalous Josephson Effect [17,40-42].

**Key experimental techniques for topological superconductors:**

- **Physical Property Measurements System (PPMS):**

PPMS is a primary tool to carry out transport, magnetic and thermal measurements of sample under a well-controlled environment. PPMS is used to determine presence of superconductivity in a candidate of topological superconductivity. Superconducting materials show a transition in temperature dependent study of resistance, magnetization and heat capacity. Among these, temperature dependent heat capacity measurements are considered to be the most reliable one to confirm bulk nature of superconductivity. Also, some indications of presence topological surface states in normal state of a topological superconductor, can be checked by performing normal state magneto-transport measurements. A conventional four-probe method is used for transport measurements while, a five-probe geometry is used for Hall measurements. Hall measurements provides information about the charge carries whether they are p-type or n-type. Hall measurements also shows how, charge carrier density changes with doping in the sample.

- **Angle Resolved Photoelectron Spectroscopy (ARPES):**

Angle resolved photoelectron spectroscopy (ARPES) is the most important tool to visualize electronic bands of the topological materials [43]. ARPES measurements provide the direct evidence of topological non-trivial character of materials and presence of topological surface states. ARPES measurement requires ultra-high vacuum (UHV) conditions inside the sample chamber. The basic principle of ARPES is based on the photo-electric effect, which shows that the electrons absorb the photons falling on the sample. If the energy of photons is greater than the work-function of the material, then the electrons get ejected out of the material. These ejected out electrons are known as photo-electrons. These photo-electrons are collected inside the electron analyzer and studied in context of their kinetic energy and emission angle. The generated photo-electrons have a direct relation with energy and momentum of electrons residing in the sample, and gives the information about energy states of the studied sample. The data obtained from ARPES measurements is analyzed by applying different models to visualize electron band structure [44]. The subsequent sections of this review explain the relevance of ARPES in different TSC.

- **Point Contact Spectroscopy:**

Point contact spectroscopy is a very useful technique to study the interaction of electrons with elemental excitations. There exist different elemental excitations in different materials as there are phonons in metals, magnons in magnets etc. In point contact spectroscopy, the spectra of these elementary excitations are measured with the help of very tiny contacts [45]. The ratio of contact diameter (d) with mean free path (l) of the samples determines the properties of electrical transports through point contacts. Accordingly, the electrical transport lies in three different regimes: (a) If the contact diameter is smaller than mean free path, and there is no scattering in the contact region, the electrical transport is said to be in ballistic regime. (b) If the contact diameter is nearly equal to the mean free path, the transport is said to be in diffusive regime. (c) If the contact diameter is greater than the mean free path, there is an inelastic scattering in the contact region, and the transport is said to be in thermal regime.

In a special case, when one of the electrodes in point contacts is superconducting, forming a normal-superconductor (N-S) interface and the transport lies in ballistic regime, point contact spectroscopy is known as point contact Andreev reflection (PCAR) spectroscopy. PCAR spectroscopy is based on the principle of Andreev reflection [45,46]. In PCAR spectroscopy, when the energy of electrons is less than the superconducting gap ($\Delta$), the conductance is enhanced by a factor of 2 causing a non-linear I-V characteristics. This non-linear I-V characteristics give information about the interaction of electrons with elementary excitations. PCAR is considered to be a handy tool to obtain some glimpses of Majorana fermions. In materials, Majorana fermions exists in form of quasi-particle excitations and the interaction of these excitations with electrons leads to occurrence of ZBCP [30-32] and the significance of ZBCP in context of Majorana fermions is discussed above. This establishes PCAR spectroscopy as an essential characterization tool for topological superconductors.

- **Muon-spin rotation (μ-SR) Spectroscopy:**

This technique involves the implantation of muons into the material. Muons are the elementary particles which has two types of charges viz. positive muons and negative muons. Muons are fermions with non-zero spin and have a very lower mass. The spin angular momentum associated with muons makes them very sensitive to the magnetic fields. For μSR, positive muons (the best to use) are implanted in the material. These muons come to rest at a stopping site by combating with the local electrostatic field into the material. Also, their lifetime is independent of materials property, while for negative muons, it depends on the material as negative muon impinges into the nucleus of local atoms.

Muons that are used in μSR facilities are generated by striking a high energy proton beam onto a graphite target. In this process, pions are produced which further decays into muons and antineutrinos, thus producing a muon beam which is spin-polarized due to parity violation. If they are present in a magnetic field, **B,** at the stopping site of the muon, the spin of muon starts to precess at Larmor frequency which is given by $\omega_\mu = \gamma_\mu \mathbf{B}$ where $\gamma_\mu/2\pi$. It is a gyromagnetic ratio which is equal to 135.5 MHz T$^{-1}$. So, by calculating the precession frequency of muons, the local magnetic field around muons can be determined. Thereby, through muons, various properties of superconductors and magnetic materials can be determined, as in absence of an external magnetic field i.e. zero fields (ZF-μSR). This technique probes very minute internal magnetic fields which are of the order of 0.1G. Thus, it helps to determine whether the TRS is broken or not in the superconducting state. When TRS is broken, there is a spontaneous rise in the magnetic field. It can be seen in terms of increased relaxation rate in the asymmetry time spectra [47-49]. Moreover, the transverse field of the μSR (TF-μSR) is used to determine the pairing symmetry in the superconducting state by calculating various parameters such as coherence length and penetration depth. In TF-μSR, a transverse magnetic field is applied in the direction perpendicular to muons spin direction to determine the mentioned parameters. The dependence of penetration depth on temperature gives the information about the superconducting gap whether it is nodal or node less.

In the next segment of this review, possible candidates of topological superconductivity are discussed. This segment consists of the results of different experiments performed on possible topological superconductors, signifying their candidature of topological superconductivity. Results of μ-SR measurements are clubbed together in a separate section to categorize candidates of topological superconductors as conventional or unconventional superconductors. This also gives the information about TRS, whether it is preserved or not in these superconductors.

**Possible Candidates of Topological Superconductivity:**

Topological Superconductivity is an emerging field of research. First of all, topological superconductivity was theoretically proposed at the interface of the topological insulator and an s wave superconductor [26]. But, later on, some glimpses of topological superconductivity were found in some unconventional and non-centrosymmetric superconductors in their intrinsic form [28,29]. Also, superconductivity can be induced in some topological materials through specific carriers doping [22] and by applying pressure [25]. Also, topological superconductivity can be observed in heterostructures that consist of TI and a superconductor or high SOC materials and superconductors [26]. In this review, we start with the Bulk Topological Superconductors candidates which consist of materials that are supposed to shows topological superconductivity in intrinsic form and topological materials that can be made superconducting through doping and pressure. We aim to cover all the bulk topological superconductors. Further, a brief discussion on heterostructures is provided, in which topological superconductivity can be probed at their interfaces.

### A. Topological Superconductivity in Bulk Materials

**A1. Intrinsic Topological Superconductors:**

- **Unconventional Superconductors:**

It is predicted that to be a candidate of topological superconductor one compound must show a full superconducting gap with odd symmetry pairing [16] but most of the known superconductors are s-wave superconductors (conventional superconductors). These superconductors have spin-singlet pairing and are mostly topologically trivial (Although, there is some theoretical evidences that in some s wave superconductors, topological superconductivity can be realized [50,51]). So, to realize topological superconductivity in the intrinsic form, one should look for the superconductors that show topologically non-trivial character, and have unconventional pairing symmetry. Unconventional superconductivity can be found in spin-triplet superconductors, as the Cooper pairs in these superconductors have spin angular momentum $S = 1$. This spin-triplet can only be paired with orbital angular momentum $L = 1,3,5....$ i.e., p, d... orbitals. The difference between spin-singlet pairing and spin-triplet pairing is that in spin-singlet pairing, the formation of Cooper pairs reduces the paramagnetic susceptibility, while in spin-triplet pairing paramagnetic susceptibility is maintained. According to pairing symmetry, the unconventional superconductors can be classified as p-wave and d-wave superconductors. It is predicted that vortex core of p-wave superconductors can host Majorana fermions [28,29]. This typical class of superconductors can be termed as possible candidates of topological superconductors. The main issue for this type of superconductor is that, their pairing can be easily broken by a magnetic field. The most studied example of such superconductors is $Sr_2RuO_4$. In this section, we'll consider about the possibility of unconventional superconductors of being a topological superconductor.

$Sr_2RuO_4$ was discovered in 1994 in an attempt to replace copper (Cu) with a 3D transition element in cuprates. The critical temperature ($T_c$) of this superconductor is very low about 1 K. The electronic structure of $Sr_2RuO_4$ is different from copper oxide superconductors as the 3d electrons in copper oxide superconductors are replaced by 4d electrons of Ruthenium (Ru) [52]. The spin state of the Cooper pairs can be determined by the Knight Shift obtained from nuclear magnetic resonance (NMR) spectroscopy. Ishida et.al. [19] in 1998 studied NMR spectroscopy of $Sr_2RuO_4$. It was found that the Knight shift remained unchanged at the critical temperature which gives evidence of the spin-triplet pairing of Cooper pair in this superconductor. Further, it was found that these results were an experimental error and recent NMR measurements suggest that the Knight shift changes at the critical temperature [53]. This suggests that $Sr_2RuO_4$ is not a triplet superconductor [53,54]. Signatures of the internal magnetic field were observed in μ-SR spectroscopy performed below superconducting transition temperature ($T_c$), which suggests that TRS is broken in the superconducting state [20,55]. Although there are clear hints of evidence that make $Sr_2RuO_4$ a suitable candidate to realize topological superconductivity that has not been confirmed experimentally. But still, there are several unanswered questions such as the weak spin-orbit coupling (SOC). Due to this weaker SOC, Majorana fermions are expected to be absent from surface states [17,56].

$Au_2Pb$ [57] and $TaIrTe_4$ [58] are other two materials that show unconventional superconductivity and can be regarded as potential candidates of topological superconductivity. $Au_2Pb$ shows superconductivity at 1.3K [57], while the unconventional nature of superconductivity is detected through low temperature STS measurements and quasi linear dependency of upper critical field on temperature [57]. $Au_2Pb$ is reported to have Dirac cone like band structure and summarized as Dirac Semi-metal (DSM) in theoretical calculations as well as in ARPES measurements [59,60]. Moreover, TaIrTe$_4$ is categorized as type-II Weyl semimetal (WSM) in both theoretical and experimental studies [61,62]. TaIrTe$_4$ shows surface superconductivity at around 1.4K [58], however it is worth mentioning that bulk superconductivity evidences in TaIrTe$_4$ are still missing. Low temperature STS measurements performed on TaIrTe$_4$ suggest that the observed surface superconductivity is unconventional in nature [58].

- **Non-centrosymmetric Superconductors:**

2-D non-centrosymmetric superconductors (NCSs) are considered to be possible candidates to show topological superconductivity. The non-centrosymmetric superconductors do not have any inversion center because of their broken inversion symmetry. These superconductors show Rashba and Dresselhaus spin-orbit couplings due to the absence of inversion symmetry. Moreover, the pairing symmetry of these systems is also unique as it shows the mixing of both s-wave pairing (spin-singlet) and p-wave pairing (spin-triplet). The topological property of these materials depends on this mixed wave pairing. If the superconducting gap associated with p wave pairing is larger than that for s wave pairing then only these materials show topological properties [63,64].

BiPd is a known NCS having a monoclinic structure with space group P2$_1$ [65,66]. It is a fully gapped superconductor having mixed pairing symmetry [67,68]. A large SOC in BiPd is a consequence of the presence of a heavy element Bi. The absence of inversion symmetry leads to splitting of bands and forms topologically non-trivial surface states in normal state of BiPd [68,69]. These surface states with Dirac nodes are observed in ARPES as well as spin-resolved ARPES measurements [70]. The nodes are found to be robust against temperature too. Dirac nodes observed in BiPd are not degenerate due to the absence of inversion symmetry [68]. Differential conductance spectra show a strong zero bias peak at the center of the vortex core of BiPd in STS measurements in field as shown in Fig. 3 which can be attributed to Caroli-Matricon-de Gennes states [68] and may host Majorana zero modes. ZBCP is observed in Point Contact Spectroscopy, and it does not split by the application of the magnetic field, this feature of ZBCP declines all reasons of its origin other than the presence of Andreev Bound States [67]. The presence of ZBCP also suggests the triplet pairing symmetry and unconventional superconductivity in BiPd.

Other non-centrosymmetric superconductors such as CePt$_3$Si, Li$_2$(Pd$_{1-x}$Pt$_x$)$_3$B show mixed pairing symmetry and can host non-trivial topological states [71,72]. Spin-orbit coupling in these materials depends on the mixing of the order parameter. Strong spin-orbit coupling in these materials is the result of higher p-wave pairing in comparison to s-wave pairing [73]. This can be confirmed by the observed line node in theoretical calculations of the superconducting gap of CePt$_3$Si and Li$_2$Pt$_3$B. The superconducting gap changes its sign with higher p wave pairing as it is given by $\Delta = \Delta s - \Delta p \, Sin \, \vartheta$ ($\Delta s$ is a superconducting gap of s-wave component and $\Delta p$ is a superconducting gap of p-wave component). The superconductors with the higher p-wave pairing, fall in the class of unconventional superconductors [74,75]. The line node is absent in the counterpart of Li$_2$Pt$_3$B in Li$_2$(Pd$_{1-x}$Pt$_x$)$_3$B system i.e. Li$_2$Pd$_3$B [75], which shows that the p-wave component is higher in Pt rich compositions, which is further responsible for higher spin-orbit coupling. These nodal superconductors with high spin-orbit coupling are supposed to host topologically non-trivial surface states [76]. All these interesting features of non-centrosymmetric superconductors with broken inversion symmetry increase their potential to be the candidates of topological superconductivity. As per our knowledge, STS measurements are still missing for these compounds which could provide evidence of the presence of Majorana fermions in these superconductors.

In the category of non-centrosymmetric superconductors, PbTaX$_2$ (X = Se, S) also grabs significant attention as these materials are superconducting with non-trivial topological surface states [77-79]. PbTaSe$_2$ is a topological nodal line semimetal (NLSM) that shows bulk superconductivity at 3.8K [77]. The bulk nature of superconductivity is confirmed by specific heat measurements [77]. Topological properties of PbTaSe$_2$ are confirmed by both DFT and ARPES measurements [80,81] which shows the presence of nodal line states near the Fermi level. These nodal line states are protected by reflection symmetry generated from Ta atomic plane. Further, PbTaSe$_2$ shows linear magneto-resistance (MR) up to 2T [Fig.4] [77] which also indicates suggestions of a non-trivial band structure. Another member of this category is PbTaS$_2$, but the stable structure of PbTaS$_2$ is not confirmed yet as it is treated as both, centrosymmetric and non-centrosymmetric. Superconductivity in centrosymmetric PbTaS$_2$ has been known from a long time [82] with bulk superconductivity at 2.6K [79]. While non-centrosymmetric PbTaS$_2$ is not yet realized experimentally. Although, there are some theoretical calculations which show that non-centrosymmetric PbTaS$_2$ is a nodal line semimetal [79]. The problem with PbTaS$_2$ is that the nodal lines are protected by inversion symmetry and time-reversal symmetry without SOC, but with the inclusion of SOC, these symmetries still show band crossing. To avoid this band crossing, an additional mirror reflection symmetry is required [79,83] which can be attained in the non-centrosymmetric system. So, it is important to realize stable non-centrosymmetric PbTaS$_2$ to make a clearer understanding of this superconducting nodal line topological semimetal.

- **Centrosymmetric superconductors:**

Apart from non-centrosymmetric superconductors some of centrosymmetric superconductors such as Bi$_2$Pd [84-86] and Sb$_2$Pd [87] show topological non-trivial character. Bi$_2$Pd crystallizes in tetragonal crystal structure and found to show superconductivity at 5.3K [84]. ARPES spectra of Bi$_2$Pd single crystal shows Dirac cone like dispersion around Γ point [84], which confirms that Bi$_2$Pd contain topological non-trivial band structure. Also, potassium (K) doped Bi$_2$Pd shows presence of surface states through surface transport measurements [86]. Another material in this category is Sb$_2$Pd. Sb$_2$Pd exhibits superconductivity at 1.35K [87]. DFT calculations and ARPES measurements of Sb$_2$Pd show the presence of a topological non-trivial electronic band structure which falls in the category of a Dirac type-II semimetal [88-90]. Sb$_2$Pd shows large magnetoresistance upto 174% at 2K (normal state of Sb$_2$Pd) under a magnetic field of 14T [91]. This also signifies the presence of surface states in the same material. The presence of topological non-trivial surface states is evident from the observed Berry phase in Sb$_2$Pd which is reported to be very close to π [91]. All these results establish both Bi$_2$Pd and Sb$_2$Pd as possible candidates of topological superconductivity.

Other examples of centrosymmetric superconductors, which are supposed to show topological superconductivity, are transition metal carbides (TMCs) viz. NbC and TaC [92-97]. Both NbC and TaC show superconductivity at relatively higher temperature which are 11.5K and 10.5K respectively [92-97]. These materials are found to have a type-II Dirac semimetal like band structure, which has been confirmed through both DFT calculations and ARPES measurements [93-97]. NbC has a NaCl type crystal structure and its superconducting temperature strongly depend on the lattice parameters [98,99] which increases or decreases with stoichiometry of carbon [100]. A powder XRD pattern fitted with F m -3 m space group symmetry is shown in Fig. 5(a). The right-hand side inset of Fig. 5(a) is showing Field cooled (FC) and Zero Field Cooled (ZFC) measurements results of NbC and the left-hand side inset shows the variation in T$_c$ with lattice parameters. The bulk electronic band structure of NbC is shown in Fig. 5(b) which shows the band inversion with inclusion of SOC. Also, calculation of Z2 invariants for the NbC system, suggest strong topology in the system. Fig. 5(c) shows how the Wannier charge centers evolve in k-space of NbC. The reported values of Z2 invariants in ref. 97 are: (i) k1=0.0, k2–k3 plane: Z2=1, (ii) k1=0.5, k2–k3 plane: Z2=1, (iii) k2=0.0, k1–k3 plane: Z2=0, (iv) k2=0.5, k1–k3 plane: Z2=1, (v) k3=0.0, k1–k2 plane: Z2=0, (vi) k3=0.5, k1–k2 plane: Z2=1. Apart from NbC and TaC some other transition metal carbides viz. CrC and VC are also predicted to show superconductivity with topological non-trivial character [101]. But there is no experimental report available in the literature to confirm this. Apart from TMCs, another centro-symmetric compound is $SnTaS_2$, which is recently found to be a nodal line semimetal with superconductivity below 3K [102,103]. $SnTaS_2$ is isoelectronic to $PbTaSe_2$ which is also a nodal line semimetal [77-81] as discussed in earlier, but both of these materials have different structures. $PbTaSe_2$ has a non-centrosymmetric structure while $SnTaS_2$ has a centrosymmetric structure [77,78,102]. There is an open scope for more experimental work regarding possible topological superconductivity in $SnTaS_2$.

**A2. Doped Topological Superconductors:**

Recently, several methods have been developed to induce superconductivity in TIs. Intercalation of the transition elements in a proper ratio is one of the prominent methods to induce superconductivity in TIs. It was confirmed by the intercalation of Copper (Cu) in $Bi_2Se_3$ [104]. After the discovery of superconductivity in Cu intercalated $Bi_2Se_3$, more transition metals such as niobium (Nb) and strontium (Sr) were found to induce superconductivity in $Bi_2Se_3$ [105-108]. Apart from intercalation in $Bi_2Se_3$, more TIs were found to show superconductivity such as $Bi_2Te_3$ and SnTe. $Bi_2Te_3$ shows superconductivity with the intercalation of palladium (Pd) [109,110]. While, SnTe shows superconductivity by doping of indium (In) [111].

- **Superconductivity doped in $Bi_2Se_3$:**

Until now three elements (Cu, Sr, Nb) are found to induce superconductivity in $Bi_2Se_3$, in this section, we review the properties of these candidate of topological superconductors studied by different groups, that are as follows:

> **$Cu_xBi_2Se_3$**

$Bi_2Se_3$ was found to be superconducting by intercalation of Cu in which, the amount of Cu ranges from 0.12-0.6. The critical temperature ($T_c$) of $Cu_xBi_2Se_3$ is approximately 3.8K for x= 0.15 [104,112]. There are two ways to add the Cu atoms i.e., either it can be intercalated between two quintuple layers of $Bi_2Se_3$ or it can replace the Bi atom. However, superconductivity is only observed when Cu is intercalated between $Bi_2Se_3$ quintuple layers. It is seen that when Cu replaces Bi atom i.e., in $Cu_{0.1}Bi_{1.9}Se_3$, it adds two holes to the system, and when Cu lies in the Vander walls gap, it adds an extra electron to the system. Thereby, the Fermi surface becomes enlarged and the surface conduction band is reshaped by the addition of electrons [113], which reduces the electron velocity. Generally, the velocity of electrons is determined by slope of the surface state band, which eventually decreases by Cu intercalation. This decrement in slope of surface states band, increases the separation between bulk and surface conduction band [113]. It shows that topologically protected surface states are more stabilized and also the topological behavior of superconductivity [113]. The addition of electrons makes Cu intercalated $Bi_2Se_3$ to be n-type and increases carrier concentration by one order higher than pure $Bi_2Se_3$[104].

Magnetization studies by Y.S. Hor et.al.[104] shows that the superconducting volume is very less in Cu intercalated $Bi_2Se_3$ which is nearly 20% as observed in zero field cooled (ZFC) measurement. But there was no sign of diamagnetism in $Cu_{0.1}Bi_{1.9}Se_3$. Superconductivity in these intercalated samples also depends upon Se stoichiometry, as a decrement in the Se ratio in samples results in suppression of superconductivity [104,114]. In resistivity vs temperature (RT) measurements, it was observed that resistance does not drop to zero as in a normal superconductor. This is due to the formation of the $Cu_xBi_{2-x}Se_3$ phase which cannot be controlled. This forms weak links in the sample which results in lower current density and lack of continuous superconducting path. Superconducting volume of $Cu_xBi_2Se_3$ can be enhanced by a significant amount by applying new growth method which is based on electrochemistry [114]. In this method, firstly $Bi_2Se_3$ single crystal was grown by a conventional melt growth method and then this crystal is wound by copper wire and another copper electrode is taken which acts as a reference electrode. The whole system is placed in an electrochemical solution. By passing a current between two electrodes, a desirable concentration of copper can be achieved. By this method, the superconducting volume can be increased up to 45%. In simple melt growth process, the quenching of the sample also plays an important role as it is essential to get the desired properties [115]. It has been observed that the normal furnace cooled sample is non-superconducting. The reason is that on slow cooling, the sample is decomposed into two-phases; one is just similar to the $Bi_2Se_3$ phase while the other is Cu rich phase. This makes the sample inhomogeneous. So, quenching at a temperature where Cu is intercalated homogenously, is an essential thing to get superconductivity [115]. It is evident from the superconducting transition observed in specific heat measurement of $Cu_xBi_2Se_3$ [Fig.6] that this compound shows bulk superconductivity despite low superconducting volume [112].

The application of pressure on superconducting materials has a large impact on their superconducting properties. It has been observed in several studies [116-118] that the application of pressure induces superconductivity in materials that do not show superconductivity at ambient pressure. While some materials show superconductivity at ambient pressure and loss its superconducting nature by applying pressure [119]. In the case of $Cu_xBi_2Se_3$, superconductivity is suppressed by the application of pressure [120]. This is because the application of pressure reduces the number of charge carriers which results in the decrease of critical temperature. Also, Cu intercalated $Bi_2Se_3$ shows reduction in the magnetoresistance by a significant amount as compared to the large magnetoresistance of $Bi_2Se_3$. It is because, the Cu intercalation increases sample inhomogeneity thereby increasing the electron scattering [121].

The nature of superconductivity in $Cu_xBi_2Se_3$ is hard to decide whether it is conventional superconductivity or unconventional superconductivity. There are evidences for both types of superconductivity in $Cu_xBi_2Se_3$, as the zero-bias conductance peak in point contact spectroscopy shows the unconventional type of superconductivity while the conduction spectrum obtained from scanning tunneling microscopy shows the conventional type of superconductivity [122,123]. The reason for a zero bias conduction peak observed in point contact spectroscopy is the existence of the Andreev Bound State (ABS) in the sample. These ABSs can host Majorana fermions as the observed ZBCP shows odd parity bulk superconducting state. Knight Shift in $Cu_xBi_2Se_3$ was measured below $T_c$ through NMR (considered to be a great tool to determine the pairing symmetry in superconductors) and it is found to remain unchanged [124]. It shows that the spin rotation symmetry is broken in the superconducting state which indicates the spin-triplet pairing of Cooper pairs in $Cu_xBi_2Se_3$ [124]. ARPES measurements made on $Cu_xBi_2Se_3$ clearly shows surface states with a Dirac point. It represents that the topological nature of $Bi_2Se_3$ remain intact after Cu intercalation [125]. All given arguments are sufficient to state that $Cu_xBi_2Se_3$ is a good choice of material as a candidate of topological superconductivity as well as for seeking the realization of Majorana fermions.

➢ **$Sr_xBi_2Se_3$**

Doping undoubtedly is a great method to induce superconductivity in a topological material. Just after the discovery of superconductivity in $Cu_xBi_2Se_3$, search for other elements was started that can induce superconductivity in $Bi_2Se_3$. Another element that induces superconductivity in $Bi_2Se_3$ is strontium. It was found that strontium (Sr), when intercalated in $Bi_2Se_3$, shows superconductivity at a critical temperature of about 2.5K for a Sr concentration of 0.05 [107]. Similar to Cu doping and intercalation, the same difference can be made in Sr doped and Sr intercalated $Bi_2Se_3$ by their powder XRD data. The c parameter of Sr doped $Bi_2Se_3$ is smaller than pure $Bi_2Se_3$ while it is greater in the case of Sr intercalated $Bi_2Se_3$ [107]. No superconducting transition is reported in Sr doped $Bi_2Se_3$ i.e., $Sr_xBi_{2-x}Se_3$ down to 2K. This is just similar to the findings in $Cu_xBi_{2-x}Se_3$ in a way that Sr doping on Bi site acts as a hole donor, while the intercalated Sr atom donates an electron and increases the charge carrier density. Electrons are effective charge carriers in $Sr_xBi_2Se_3$ as confirmed by the negative slope of resistivity vs magnetic field plot found in Hall measurement [107]. The way the Hall coefficient depends on the temperature in the Sr doped system clarifies the increase of charge carrier density by lowering the temperature [107]. Now, here the question is that whether these intercalated atoms reside in Vander Waals gaps between two quintuple layers as in the case of $Cu_xBi_2Se_3$ or they reside elsewhere. The answer was found in a study made by Zhuojun Li et al. [126] that Sr atom atoms are doped in $Bi_2Se_3$ at interstitial positions as well as in Vander Waals gap. The interstitial site of Sr atoms is found to be energetically metastable and responsible for the emerged superconductivity [126]. Superconductivity in $Sr_xBi_2Se_3$ also depends on the cooling process during crystal growth, as it was found that carrier density is higher for ice quenched samples compared to normal furnace cooled sample for optimal Sr concentration, which is responsible for superconductivity. It is also observed that the normal furnace cooled sample does not show zero resistivity and the observed diamagnetism in these samples is also weaker than that of the quenched sample [126]. A large superconducting volume up to 91% can be achieved by changing the amount of Sr which is much greater than that was achieved in $Cu_xBi_2Se_3$. Also, its greater stability in the air for several days makes $Sr_xBi_2Se_3$ a better choice of material as a candidate of topological superconductivity. [107,127]. The existence of surface states in $Sr_xBi_2Se_3$ is evident from the observed Shubnikov−de Hass oscillations in resistivity vs magnetic field curves [107]. An ARPES study of $Sr_xBi_2Se_3$ shows an observable Dirac point, which indicates topological non-trivial surface states are present in the sample. This indicates that the intercalation of Sr atoms in $Bi_2Se_3$ does not affect the surface states [128]. When the ARPES data of $Sr_xBi_2Se_3$ is compared with that of $Bi_2Se_3$, it was found that the Dirac point moves downward below Fermi level towards the higher binding energy. The downward movement of Dirac point shows the increment in charge carrier concentration. The shift in Dirac point is found to be more for $Cu_xBi_2Se_3$ as compared to $Sr_xBi_2Se_3$ as shown in [Fig.7(a), (b)]. This shows that Cu intercalation increases the charge carrier density more than Sr [128,129] intercalation. In a time-resolved ARPES study (TRARPES), the surface of $Sr_xBi_2Se_3$ was found to host topological surface states and 2D quantum well states simultaneously. By analyzing different scattering mechanisms, it was found that the excited 2D quantum well states had a shorter time period [130].

In an attempt by Guan Du et.al. [127] to analyze the nature of superconductivity in $Sr_xBi_2Se_3$, it was observed that there exist two superconducting gaps in the sample. One superconducting gap shows a s-wave pairing and the other shows an anisotropic s wave pairing symmetry of electrons. These superconducting gaps were found to exist due to two different types of electrons: s-wave superconducting gap is due to the bulk electrons while the anisotropic s-wave gap is due to the topological surface state [127]. This particular feature shows the role of surface Dirac electrons in the superconductivity of $Sr_xBi_2Se_3$. The behavior of Cooper pairs formed by surface Dirac electrons can be interesting as these can be affected by other phenomena shown by surface Dirac electrons such as spin momentum locking [127]. Magnetic transport measurements of $Sr_xBi_2Se_3$ show the angular dependency of the upper critical field and breaking of rotational symmetry, which can be attributed to unconventional spin-triplet superconductivity. Moreover, these findings suggest that $Sr_xBi_2Se_3$ can be regarded as a nematic superconductor exhibiting two-fold symmetry respectively. This nematicity is found to emerge from normal state above $T_c$ as observed through Angle-Resolved Specific Heat (ARSH) measurement at different magnetic fields [131-135]. A high anisotropic magnetoresistance is also observed for $Sr_xBi_2Se_3$, which leads to the presence of chiral anomaly features found in topological compounds [136].

> **$Nb_xBi_2Se_3$**

Another choice of an element that can induce superconductivity in $Bi_2Se_3$ is niobium (Nb). It is well established in the literature that niobium intercalation in $Bi_2Se_3$ can induce type-II superconductivity with a critical temperature in a range of 2.5K to 3.4K [105,108]. The diamagnetic transition observed for this material in ZFC measurements is very sharp as shown in Fig. 8, which can be regarded as evidence of bulk superconductivity. The superconducting volume fraction of $Nb_xBi_2Se_3$ is quite large compared to $Cu_xBi_2Se_3$ and is found to be nearly 100% [105]. ARPES measurement of $Nb_xBi_2Se_3$ clearly shows the Dirac point, which indicates the existence of topological non-trivial surface states [105]. Nb intercalation in $Bi_2Se_3$ incapacitates more charge carriers which lead to the occurrence of superconductivity as pristine $Bi_2Se_3$ does not have enough carriers. Interestingly, superconductivity in both Sr doped and Cu doped $Bi_2Se_3$ does not depend on charge carrier density while in Nb-doped $Bi_2Se_3$ superconductivity depends on the same [107]. Interestingly, despite of the fact that $Bi_2Se_3$ is non-magnetic and Nb is slightly paramagnetic, $Nb_xBi_2Se_3$ is found to be magnetic in nature. This observed magnetic behavior of $Nb_xbi_2Se_3$ can be attributed to its triple pairing symmetry [105]. This type of pairing induces an internal magnetic field in the sample resulting in broken rotational symmetry. Torque magnetometry measurements done by Tomoya Asaba et al. [137] confirm the breaking of rotational symmetry in $Nb_xBi_2Se_3$. The presence of robust zero bias conduction peak with sharp dips in $Nb_xBi_2Se_3$ also indicates unconventional p wave superconductivity [138,139] in $Nb_xBi_2Se_3$.

The large superconducting volume of $Nb_xBi_2Se_3$ as compared to $Cu_xBi_2Se_3$ makes it a good choice of material as a candidate of topological superconductivity. Also, $Nb_xBi_2Se_3$ show bulk superconductivity transition in specific heat measurements, which signifies that $Nb_xBi_2Se_3$ is a bulk superconductor [140]. This bulk superconducting transition is interesting in itself as there are very few reports showing superconducting transition in heat capacity measurements of superconducting TIs. Single crystal of $Nb_xBi_2Se_3$ can be grown by a conventional melt growth method following the heat treatment which is optimized by our group as shown in Fig. 9(a) [141]. Nb intercalation in the Vander Waals gap of $Bi_2Se_3$ depends on heat treatment. As a structural study made on crystals grown by melting at temperature $800°C$, shows the presence of two phases one $Nb_xBi_2Se_3$ and the other one is of $NbBiSe_3$, in which Nb substitute bismuth on Bi site [106] [Fig. 9(b)]. This Nb substituted phase seems to be stabilized with increasing Nb content as shown in the inset of Fig. 9(b). Since the $NbBiSe_3$ phase is non-topological and superconducting at the relatively lower temperature 2.3K [142], it is more significant to choose a proper heat treatment to grow pure $Nb_xBi_2Se_3$ crystals. The crystal grown at a higher temperature at $950^0C$ under a well optimized heat treatment by seems to be in single phase [141], as no peak is observed of this phase in the XRD pattern [Fig. 9(c)]. Also, it was confirmed by Raman Spectroscopy measurements, where it was observed that Raman modes were almost similar as of $Bi_2Se_3$ except in case of the $A^2_{1g}$ mode. As, it shows slight shift due to the insertion of extra Coulombic layer in the Vander Waals gap, as shown in Fig. 9(d) [108]. It was found that to achieve superconductivity, Nb should be integrated in the Vander Waals gap, not at the site of the Bi atom. Magnetic measurements show different resistivity in the field parallel to the c- axis, and field parallel to the ab plane. This shows the anisotropic nature of $Nb_xBi_2Se_3$ [105]. The effect of pressure on the superconductivity of Nb-doped $Bi_2Se_3$ is different from that observed for Cu and Sr-doped $Bi_2Se_3$. $T_c$ was found to increase with the application of pressure untill 0.56 GPa in Nb doped samples but in the case of Sr and Cu-doped samples, the same was suppressed with the application of pressure [Fig. 10][143]. Angle dependent magnetization measurements done by B. J. Lawson et.al. [144] show quantum oscillations (dHvA effect) in $Nb_xBi_2Se_3$ with two frequencies [Fig. 11]. This indicates the existence of electronic states having more than one band in $Nb_xBi_2Se_3$ which were not observed in $Cu_xBi_2Se_3$.

An angle dependent magnetoresistance study of $Nb_xBi_2Se_3$ done by Junying Shen et al. [145], showed that the superconducting transition in the presence of a magnetic field has a strong angular dependence. This study completely discarded the possibility of three-fold symmetry in the crystal and gave an indication of two-fold symmetry and nematic superconductivity in $Nb_xBi_2Se_3$. The nematic superconductivity in $Nb_xBi_2Se3$ was also confirmed by the experiment done by M. P. Smylie et.al. [146] via penetration depth measurement [145,146]. Another interesting feature was observed in $Nb_xBi_2Se_3$, where the coexistence of superconductivity and ferromagnetism was experimentally observed by Noah F. Q. Yuan et al. [147]. The interpretation from their study is just the opposite of that observed by Tomoyo Asaba et al. [137], which suggests that superconductivity in $Nb_xBi_2Se_3$ is nematic because the rotational symmetry is spontaneously broken just like in the case of $Cu_xBi_2Se_3$. But the work of Noah F. Q. Yuan et.al. [147] indicates that the time-reversal symmetry is broken in $Nb_xBi_2Se_3$ which causes the topological phase of superconductivity in $Nb_xBi_2Se_3$ to be chiral. The ferromagnetic order observed in $Nb_xBi_2Se_3$ is due to the evolution of the chiral topological phase because a nematic phase can never induce a ferromagnetic order. This is also the reason behind the difference in superconducting properties of $Cu_xBi_2Se_3$ and $Nb_xBi_2Se_3$. Such characteristics of superconductivity depend on the type of dopant. In $Nb_xBi_2Se_3$, Nb atoms have a net magnetic moment while in $Cu_xBi_2Se_3$, Cu atoms do not have any magnetic moment which is responsible for the nematic phase in $Cu_xBi_2Se_3$.

From all above three, $Sr_xBi_2Se_3$ stands out a better candidate material for topological superconductivity as it has a high shielding fraction and it was found to be robust against environmental impurities. Another reason for this material to be superior to other candidates is that it is easier to grow in single-phase while, $Cu_xBi_2Se_3$ and $Nb_xBi_2Se_3$ are found to have some amount of secondary phases.

- **Superconductivity in doped $Bi_2Te_3$:**

$Bi_2Te_3$ was found to be superconducting with doping of elements like palladium (Pd) and thallium (Tl) [109,110]. $Pd_xBi_2Te_3$ was found to be superconduct with the critical temperature near 5.5K for x=1[148]. The superconducting fraction found in this superconductor is very low which is approximately 1%. This is due to the inhomogeneity of the sample. $Bi_2Te_3$ exhibits an antisite defect which prevents proper Pd intercalation in this sample. It can be reduced by doping of indium on Bi site [109,149]. More work is yet to be done to give the full description of superconductivity in this system.

Thallium doping was also found to induce superconductivity in $Bi_2Te_3$ with a critical temperature ~ 2.2K and with a shielding fraction of 95% [150]. Superconductivity in $Tl_xBi_2Se_3$ is affected by mechanical forces as it is observed that superconductivity is absent in powdered samples. Also, the superconducting volume is temperature sensitive and found to decrease even when the sample is placed at room temperature [150]. The type of carriers in $Tl_xBi_2Te_3$ is also different from that of all the above-reviewed samples, as the charge carriers in this system are of p-type [151]. The topological nature of $Tl_xBi_2Te_3$ was confirmed by the ARPES study [151]. It was also found that Tl atoms were not settled in the Vander Waals gap but were incorporated on the Bi site and the c parameter was found to decrease as compared to pure $Bi_2Te_3$ [152,153]. Further studies are yet to be done to analyze the superconductivity in these systems such as the type of superconductivity, pairing symmetry etc.

- **Superconductivity in doped SnTe:**

SnTe is a well-studied alloy for its thermoelectric and topological properties [154,155]. It is found, when a vacancy is created on Sn site in SnTe, the number of holes increases, and when carrier density increases to $10^{20}/cm^3$, SnTe becomes a superconductor with a critical temperature below 0.3K [156]. Doping with In atoms is also found to increase the number of holes in the system which increases $T_c$ up to 4.5K [156-161]. The critical temperature was found to decrease to 2.5K by the application of pressure up to 2.5GPa [161] as shown in Fig. 12. Indium doped SnTe remains topologically nontrivial, but only the spectra are broadened due to In doping which was confirmed by the ARPES study. The TRS was found to be preserved as observed in the μ-SR study [158,162]. Point contact spectroscopy study of $Sn_{1-x}In_xTe$ showed the presence of ZBCP which was suppressed in the presence of a magnetic field. This study confirmed the presence of Andreev Bound State (ABS) in $Sn_{1-x}In_xTe$ [163]. The presence of ABS in $Sn_{1-x}In_xTe$ showed the unconventional nature of its superconductivity [163]. Also, an interesting feature is observed in the nature of superconductivity in $Sn_{1-x}In_xTe$ that both conventional even pairing and unconventional odd pairing symmetries are competing with each other. In the rhombohedral phase of low $T_c$ < 1.2K, superconductivity was found to be of an unconventional type while for the cubic phase of high $T_c$ > 1.2K, the conventional even pairing dominates. This behavior happens due to scattering by a nonmagnetic impurity which is a well know phenomenology in superconductors i.e. doping of In atoms as a nonmagnetic impurity suppresses the odd pairing symmetry in them [163-165]. Also, the thermal conductivity measurements of $Sn_{1-x}In_xTe$ show the node-less superconducting gap with odd parity pairing, and the μ-SR study of the same material also indicates a fully gapped superconducting gap [158,166]. The candidature of $Sn_{1-x}In_xTe$ to be a topological superconductor for higher In doping is under question as the gap structure is not well defined. It may have any one of two states viz. topological trivial $A_{1g}$ or topological non-trivial $A_{1u}$ [166,167].

In many earlier studies, SnTe has been alloyed with different metals in which, a polycrystalline $(Pb_{1-z}Sn_z)_{1-x}In_xTe$ was found to be superconducting with max. $T_c$=4.7K [168,169] and $Pb_{1-z}Sn_zTe$ was found to be topological crystalline insulator for z>0.35[170,171]. Both of these arguments indicate the possibility of topological superconductivity in $(Pb_{1-z}Sn_z)_{1-x}In_xTe$ [172]. Scanning Tunneling Spectroscopy (STS) study of $(Pb_{1-z}Sn_z)_{1-x}In_xTe$ discards the possibility of the sample to be a topological superconductor as the superconducting gap was found to be conventional with s-wave pairing [173].

From above discussion on doped topological insulators and doped crystalline insulator, it is still unclear why the intercalation of some certain elements (Cu, Nb, Sr, Tl, Pd) in $Bi_2Se_3$ and $Bi_2Te_3$ induce superconductivity. A significant impact of this material intercalation can be seen in terms of increased carrier density which can be attributed as a possible reason for induced superconductivity but the question remains unanswered: why does not intercalation of other elements induce superconductivity in these TIs? Among these elements, all elements are superconducting in their elemental form except Cu, so it can be speculated that individual superconductivity of these elements can also plays a role in the superconductivity of doped TIs. In the case of Cu, the high metallicity or high carrier density of Cu may be the possible reason for observed superconductivity. The intercalation of such materials that are superconducting individually can be viewed as a natural heterostructure inside the lattice as an extra superconducting layer is inserted between unit cells of TIs. To explore this, more experimental work is required that can show the behavior of these TIs on the intercalation of superconducting materials. While in case of In doped SnTe, occurrence of superconductivity is well described but there are a lots of unanswered questions regarding superconducting gap and topological superconductivity.

### A3. Tin-based superconductors ($Sn_{1-x}Sb_x$, SnAs & $Sn_4As_3$):

SnTe along with its specific dopant was found to be superconducting, while there are other Sn-based systems, which showed the possibility to be the candidate of topological superconductivity. One such Sn-based alloy system is SnSb. The superconductivity of SnSb at 2.3K was found a long ago [174]. It is a type-II superconductor having n-type charge carriers [175]. Its topological nature is not explored significantly yet. In specific heat measurements, SnSb was found to be a fully gapped s-wave superconductor [172]. Antimony (Sb) was known to have topological properties [176]. In the SnSb system, Sn atoms were intercalated between Sb layers [177,178] which can serve as natural heterostructures. Heterostructures can alter the topological properties of a material [179-181]. Interestingly, when changing the composition of SnSb in $Sn_{1-x}Sb_x$, it is observed that when 'x' reaches 0.6, the XRD pattern shows some satellite peaks as shown in Fig. 13. These peaks indicate the presence of two-unit cells with different c-parameters [178], one with the same c-parameter as of SnSb and the other with nearly doubled unit cell along the c-axis. This doubled unit cell occurs due to the insertion of Sb atomic layers in the unit cell of Sn as can be seen in Fig. 13. Unit cells of $Sn_{1-x}Sb_x$ are drawn at just right of their respective XRD plots. It shows that Sb atoms make their independent atomic layer in $Sn_{0.4}Sb_{0.6}$ [178,182], making it an example of natural heterostructure. First principle calculations on the band structures of $Sn_{1-x}Sb_x$ with SOC showed the opening of the gap between the valence band and conduction band [Fig. 14 (a), (b) and (c)]. These calculations indicated the presence of topological non-trivial states in $Sn_{1-x}Sb_x$ [181,182]. Interestingly, the critical temperature of SnSb was found to increase from 2.3K as we go to $Sn_{1-x}Sb_x$ composition [178,181,182]. Following this $Sn_{0.4}Sb_{0.6}$ was found to show superconductivity up to 4.6K [182] [Fig.15] in DC magnetization measurements viz. FC and ZFC measurements. It can be questioned whether the observed superconductivity above 4K is due to Sn only or due to $Sn_{1-x}Sb_x$, but the M-H isotherms showed type-II superconductivity up to 4K [inset of Fig. 15] and confirmed $Sn_{1-x}Sb_x$ to be dominating the superconducting phase. $Sn_{1-x}Sb_x$ lacks experimental evidence that can show the topological behavior of these materials while an observable gap induced by the inclusion of SOC in DFT measurements gives a hope for possible topological states in this same. More experiments such as ARPES measurements and scanning tunneling spectroscopy measurements are still needed to check its topological behavior.

Another example of a Sn-based superconductor which is regarded as a candidate for topological superconductivity is SnAs. SnAs has NaCl type cubic structure and its critical temperature ranges from 3.5K-4K [174,183,184] [Fig. 16(a)&16(b)]. ARPES measurements of SnAs showed the splitting of bands due to spin-orbit coupling. It confirmed that the SOC has a significant impact on the band structure of SnAs [185], this is in agreement with theoretical calculations [184] as shown in Fig. 17(a). The surface states were also mapped at the Γ point in ref. 184, and shown in Fig. 17(b). Also, calculations of Z2 invariants for SnAs system suggested the presence of strong topology [184]. Fig. 17(c) and 17(d) show the Fermi surface of SnAs and the evolution of Wannier charge centers in K-space of SnAs respectively.

Superconductivity in SnAs is conventional with s-wave pairing [186]. There is also uncertainty in the nature of superconductivity of SnAs. In ref. [183], SnAs seems to be a perfect type-I superconductor while in ref. [184], it is found to be weak type-II superconductor. Fig. 16(b) shows FC and ZFC measurements of a SnAs sample, where a clear difference between FC and ZFC signals can be seen. It is showing that in FC measurements, magnetic flux has been trapped which is a clear sign of type-II superconductivity. The inset of Fig. 16 is shows an opened M-H loop of SnAs at 2K, which is a feature of type-II superconductivity. The upper critical field is nearly 5 times larger than the lower critical field which is not possible for a type-I superconductor.

Summarily, it can be stated that SnAs is an interesting system that should be explored in the context of its superconducting properties and consider its topological properties as well.

Another compound of Sn and As, that can be considered as a potential candidate of topological superconductivity is $Sn_4As_3$ [187]. $Sn_4As_3$ is a non-centrosymmetric superconductor with a $T_c$ around 1.1K [187]. Presence of surface states are evident from ARPES measurements of $Sn_4As_3$ [187]. More work is required to get more insight into the topological properties of this compound.

**A4. Iron-based Superconductors:**

Iron (Fe) based superconductors such as $FeSe_{0.5}Te_{0.5}$ and LiFeAs are found to have topological non-trivial surface states on the surface in the first principal calculations and ARPES measurements [188], while the evidence of bulk topological properties are still missing. Iron-based superconductors have always been a hot topic for research ever since from their discovery in $La[O_{1-x}F_x]FeAs$ by Kamihara et.al. in 2008 [189]. Over the seven years since their discovery, the monolayer of FeSe grown on the $SrTiO_3$ (STO) substrate showed superconductivity near to 100K [190] and this system (FeSe/STO) emerged as a model system to study high $T_c$ superconductivity. ARPES measurements of a monolayer FeSe shows that its electronic band structure is different to bulk FeSe. Basically, the hole pockets observed in bulk FeSe around Γ point. These are absent in case of monolayer FeSe due to lattice mismatch between FeSe monolayer and STO substrate [191-196]. This lattice distortion is also responsible for the evolution of the gapped phase in bulk. The associated SOC in FeSe monolayer, induces 1D topological edge states around M point which are confirmed by scanning tunneling spectroscopy and ARPES measurements [196,197] too. Hence, monolayer FeSe can be regarded as a possible candidate of topological superconductivity. While in the bulk FeSe system, topological non-trivial states are predicted in first principle calculations when Se is substituted by Te [198,199]. The doping of Te on the Se site was found to increase SOC as well as be responsible for the emergence of non-trivial edge states [200]. However, it was difficult to observe these surface states through ARPES measurements as the gap induced due to SOC was calculated to be very low near to 10 meV [201]. STS measurements performed on $FeTe_{0.55}Se_{0.45}$ showed the presence of a sharp zero biased conduction peak in the vortex core which did not split by moving apart from the vortex center and remained at zero energy [Fig. 18(a),(b)]. It gave evidence of the presence of Majorana bound states (MBS) in the vortex core [202]. Also, zero energy bound states are reported to appear in STS measurements carried out on monolayer of $FeTe_{0.55}Se_{0.45}$ [203], this also emphasize the possible candidature of $FeTe_{0.55}Se_{0.45}$ as a topological superconductor. Recently, zero energy modes are observed in one unit cell of FeSe and $FeTe_{0.5}Se_{0.5}$ deposited on STO substrate [204]. These results establish the iron arsenide system as an ineluctable system in the context of topological superconductivity. Still, this system requires more experimental evidence before it can be regarded as a topological superconductor. Some of heavily electron-doped FeSe ($Li_{0.84}Fe_{0.16}OHFeSe$) and some other arsenide materials (CaAs & $CaFeAs_2$) are also predicted to be candidates of topological superconductivity [205,206].

**A5. Transition Metal Di-Chalcogenides:**

Transition Metal Di-chalcogenides (TMDs) are interesting materials due to their 2-D nature and topological properties. Some of these TMDs having general formula $TMX_2$ (TM= Transition Metal & X= Chalcogen) such as $MoTe_2$ [207], $WTe_2$ [208], $PdTe_2$ [209], $PtTe_2$ [210], $PtSe_2$ [211], $NiTe_2$ [212] are supposed to be type-II Dirac/ Weyl semimetals. Among these materials, $MoTe_2$ and $PdTe_2$ are found to be superconducting in their intrinsic form at 0.10K and 1.64K respectively at ambient pressure [213-215]. $PdTe_2$ was also found to show superconductivity with enhanced $T_c$ by doping of Cu [216] and Au [217]. Cu doped $PdTe_2$ showed superconductivity at an enhanced $T_c$ of 2.6K [216] while Au doped $PdTe_2$ showed superconductivity at around 4.65K [217].

In a recent theoretical study, a monolayer of $NiTe_2$ was suggested to be superconducting at 5.7K and it was also predicted that its bilayer intercalated with lithium would be superconducting with a $T_c$ as high as 11.3K [218]. Superconductivity in Te deficient $NiTe_2$ at 4K was also reported by the intercalation of Titanium (Ti) in the Vander Waals gap, which was independent of pressure [219]. Similar behavior was also observed by the same group for $ZrTe_2$, in which $ZrTe_2$ was found to be superconducting by intercalation copper atoms. Also, first principal calculations made on this system indicates towards the presence of non-trivial surface states [220]. $MoTe_2$ is another TMD that is regarded as Type-II Weyl Semimetal [207] with high magnetoresistance. In its intrinsic form, $MoTe_2$ appears to be an unconventional superconductor with two band superconductivity as revealed very recently in transport measurements as well as in point contact spectroscopy in ref. 214. Its superconducting properties can be enhanced with doping of S on the Te site making it $MoTe_{2-x}S_x$, and max. $T_c$ that was observed to be 1.3K for x= 0.2 [221,222]. $IrTe_2$ is another fascinating TMD that is superconducting near to 2K and its $1T-IrTe_2$ phase is type-II

DSM [223,224]. This phase was found to be stabilized by Pt doping [223]. Intrinsic IrTe$_2$ was found to show a linear MR up to 200% under a magnetic field of ±15T [223] which was also a signature of type-II DSMs. Apart from these materials, the monolayer of some TMDs such as WS$_2$ [225,226], NbSe$_2$ [227,228] were also found to show topological superconductivity. These reports suggest TMDs are good class of materials to gain more insight about topological superconductivity and the presence of Vander Waals gap in these materials provides a platform for the intercalation of different elements to tune the physical and chemical properties.

**A6. Pressure-induced Superconductivity in Topological Insulators:**

Superconductivity in topological insulators is a result of tuning of electronic structure. This electronic structure can be tuned either by doping which we have already discussed in the above sections and another method is through the physical application of pressure. Most of the known topological insulators were found to be superconducting accompanied by a change in crystal structure by application of pressure [229-233]. Bi$_2$Se$_3$ shows a superconducting transition at 4.4K under the pressure of 12GPa [229]. The crystal structure of Bi$_2$Se$_3$ is also changed from R -3 m to C 2/m by pressure above 10GPa. Another structural transformation occurs in the same by pressure above 20GPa in which crystal structure is changed C 2/c and above 29GPa crystal structure changes to BCC structure with I m -3 m symmetry [230]. T$_c$ varies non-monotonically with pressure as T$_c$ increases to 8.2K at 17.6 GPa and then decreases to 6.5K at 23 GPa [229]. Bi$_2$Te$_3$ is another popular topological insulator that undergoes superconducting transition at a critical temperature of about 3K at a pressure of 4GPa [230]. Bi$_2$Te$_3$ also undergoes crystal structure transformation by applying pressure as its crystal structure changes from rhombohedral R-3 m to monoclinic C 2/m structure at a pressure of 8GPa with an increment in T$_c$ to 8K [231]. Sb$_2$Te$_3$ also shows superconductivity at 3K by the application of pressure above 4GPa [232]. Sb$_2$Te$_3$ undergoes two structural transformations first from R -3 m space group to C 2/m at a pressure of about 9.3GPa and the second one to I m -3 m at about 19.6 GPa [233]. Critical temperature increases from 3K to 7K at 7.5GPa and remains constant above that [232]. Table 2 shows all structural transformations and critical temperatures of topological insulators at various pressures. In another category of topological materials that can be made superconducting by the application of pressure are some Type-II Dirac/Weyl semimetals. In Weyl semimetal, WTe$_2$ [234], MoTe$_2$ [213] and TaP [235] were found to be superconducting under pressure while in the Dirac semimetal category Cd$_3$As$_2$ [236], ZrTe$_5$ [237] and HfTe$_5$ [238] show superconductivity by the application of pressure. Here, it is worth mentioning that apart from pressure induced superconductivity, Cd$_3$As$_2$ was found to show unconventional superconductivity which was induced by a point contact [239]. The structural transformation and T$_c$ at different pressures for both Weyl semimetal and Dirac Semimetal is given in Table 3 and Table 4 respectively. In this way, application of pressure is proved to be an effective method to induce superconductivity in topological materials for achieving topological superconductivity. It is difficult to realize other characteristics of topological superconductors such MZMs, ZBCP, etc. under high-pressure conditions and thereby realize the Majorana fermions. So, the doping method is comparatively much convenient to realize topological superconductivity than the application of pressure.

From the above discussions, it is clear that the superconducting order parameter and pairing symmetry plays a crucial role in considering a bulk superconductor as a candidate for topological superconductivity. Also, it is theoretically assumed that Majorana Fermions can be traced at the vortex core of unconventional p wave superconductors. Therefore, it becomes very important to precisely investigate the pairing symmetry of bulk superconductors that are supposed to be candidates of topological superconductivity. For the same materials, muon spin rotation (μSR) is the most reliable technique to probe the pairing symmetry of superconductors. This technique is also used to check whether the TRS states are broken or preserved in superconducting state. Here, the results of μ-SR studies made on above reviewed materials are discussed as a separate section to get a in depth insight about the pairing symmetry and the nature of TRS.

### B. μ-SR studies of bulk candidates of topological superconductivity:

In Sr$_2$RuO$_4$, which is a known unconventional superconductor and predicted to be a candidate of topological superconductivity, its ZF-μSR shows an exponential increment in the relaxation rate of asymmetry time spectra below T$_c$ due to a spontaneous rise of the magnetic field [20,240,241]. This spontaneous magnetic field signifies the breaking of TRS and the presence of triplet pairing symmetry. In non-centrosymmetric superconductor PbTaSe$_2$, ZF-μSR measurement shows that the relaxation rate of asymmetry time spectra does not change appreciably down to 0.025K, which is considerably below the T$_c$ of PbTaSe$_2$ [242]. This confirms that TRS in preserved in PbTaSe$_2$ in its superconducting state. It indicates that there is no spontaneous magnetic field below $T_c$ which neglects any possibility of the presence of p-wave states. It is more unlikely for Majorana Fermions to be observed in the vortex core of this superconductor although it has topological non-trivial surface states [80,81]. In centrosymmetric superconductors NbC and TaC no evidence of TRS breaking is observed in ZF-μSR. Also, the pairing symmetry of superconductivity in both NbC and TaC is s-wave pairing [94].

In the case of doped topological insulators, $Cu_xBi_2Se_3$ was the first that was found to be superconducting [104]. Its ZF-μSR measurements are not available in the literature to date due to its low superconducting volume fraction, while TF-μSR spectra show that superconductivity is fully gapped in $Cu_xBi_2Se_3$ [243]. But in other doped superconductors, such as in $Sr_xBi_2Se_3$, the TRS is broken in its superconducting state as revealed in ZF-μSR measurements [244] in which, the relaxation rate in the asymmetry time spectra is found to increase at 0.09K from the normal state i.e. at 3.6K [Fig. 19]. It suggests there is a spin-triplet pairing with breaking of TRS in this superconductor. Its TF-μSR spectra [244,245] also confirms the presence of bulk superconductivity with spontaneous magnetization below $T_c$ [Fig. 20], suggesting spin triplet pairing with broken TRS in $Sr_xBi_2Se_3$. In $Nb_xBi_2Se_3$ which is also a doped topological superconductor, ZF-μSR measurements show the absence of any spontaneous magnetic field below $T_c$ down to 0.4K [140]. It signifies that TRS is preserved in the superconducting state. The dependence of the penetration depth on the temperature, calculated from TF-μSR spectra was found to be similar to the unconventional high $T_c$ superconductor [140] which suggests that $Nb_xBi_2Se_3$ shows unconventional superconductivity along with preserved TRS.

ZF-μSR spectra of $In_xSn_{1-x}Te$ shows that the relaxation rate in asymmetric time spectra is found to be constant at a temperature considerably below to $T_c$ [158]. It suggests that TRS is not broken in the superconducting state and there is no component of unconventional p-wave pairing present. ZF-μSR confirms the superconductivity to be fully gapped with s-wave pairing. This also reduces the possibility that Majorana Fermions can be detected in vortex core of this material. In TMDs, $MoTe_2$ and $PdTe_2$ are type-II Dirac Semimetals and also have superconducting properties in the intrinsic form [207,209]. ZF-μSR spectra of $MoTe_2$ shows that TRS is preserved below $T_c$ in both conditions, that is ambient and under pressure. It also neglects any possibility of triplet pairing, which has also been confirmed in TF-μSR [246]. Here, the superconductivity was found to have s-wave pairing with 2 superconducting gaps. Also, penetration depth and temperature dependence showed the superconductivity to be nodal [246]. The TF-µSR spectra of $PdTe_2$ suggested that there was a mixing of type-I and type-II superconductivity [247] and ZF-µSR spectra suggested that $PdTe_2$ was a TRS invariant superconductor with conventional s-wave pairing [248]. Here we summarized all above-mentioned candidates of topological superconductivity according to breaking and preserving of TRS in a superconducting state. In Table-5, these superconductors are given with their status of TRS in the superconducting state, type of superconducting pairing symmetry and type of topological material.

### C. Topological Superconductivity at Material Interfaces

Until now, we have discussed only the materials that show topological superconductivity in bulk form. But the superconductivity can be induced at the different interfaces such as an interface between two materials of different classes such as semiconducting and superconducting interfaces [249-252], between TIs and superconductor [253-259] and between the magnetic material and superconductor [260,261]. In this section, we will discuss all these observed interfaces of topological superconductivity.

**C1. Topological Superconductivity at Semiconducting and Superconducting Interfaces:**

The interface between the superconductor and semiconducting material is considered to be the most common approach to realize topological superconductivity. The most interesting consideration about these interfaces is that none of these two materials forming an interface is topologically non-trivial. It is well known that when an interface is formed the energy bands of both materials changes appreciably. It has been predicted earlier in theoretical studies that a combination of a s-wave superconducting pair and a high SOC can be driven into a topologically superconducting state by the application of a magnetic field [262-264]. The first experimental observation of topological superconductivity at the superconductor-semiconductor interface was reported by Mourik et al. [249] in which they made measurements on InSb nanowires contacted with two electrodes one Au and other NbTiN (superconducting).

First, it was shown by theoretical calculations by Sau et.al. [27] that a heterostructure consisting of a semiconducting layer of high SOC sandwiched between a s-wave superconductor and a ferromagnetic insulator [Fig. 21(a)] can host Majorana Fermions or show topological superconductivity. Here a ferromagnetic insulator produces a Zeeman field in a direction perpendicular to the semiconductor, which forms a gap between two spin-orbit split bands. The proximity effect generates a weak superconducting field, which depending on the position of Fermi level. This superconducting field starts pairing of electrons of opposite momenta and opposite spin which forms a gap denoted as Δ. The condition that this superconducting phase would be topological is given by

$$E_z > (\mu^2 + \Delta^2)^{1/2} \quad (1)$$

where μ is Fermi level and $E_z$ is Zeeman Energy. Zeeman energy is given by $E_z = g\mu_B B/2$ where g is Lande g factor and $\mu_B$ is Bohr Magnetron. To achieve the topological phase condition, the value of Lande g factor should be high-that is why semiconductors with high SOC are used. After that, a more simplified arrangement was proposed in ref. 265, in which a perpendicular magnetic field is applied to a semiconductor instead of using a ferromagnetic insulator. This arrangement provides a tunable Zeeman field. This can be understood with the help of Fig. 21(b) which consists of an E-K diagram of spin up and spin down bands of a semiconducting material with high SOC. Extraordinary SOC present in these materials creates a magnetic field $B_{so}$ which is given as **$B_{so}$ = p\*E** (where p and E are momentum and electric field which are perpendicular to each other). This field separates electrons according to their spin states as shown in Fig. 21(b). There are two bands corresponding to different spin states viz. spin up and spin down. In the absence of a magnetic field, these bands cut each other at p=0. Chemical potential μ is considered as the relative position of Fermi level from this crossing point of two bands. When we apply a magnetic field, B, which is perpendicular to $B_{so}$, these spin states are mixed, and a gap is induced which is equal to Zeeman energy. This gap is known as Zeeman gap. This Zeeman gap tends to modify in proximity of a superconductor, and the two spin states (up and down) with opposite momenta starts pairing due to superconducting field (magnetic field emanating from superconducting current) of the superconducting layer. Modification of Zeeman field and pairing of spin sates result in induction of a superconducting gap at $p = p_f$ [4,266].

This approach was first used by Mourik et.al. [249], where they used InSb semiconducting nanowires. They used InSb because of its high SOC character. These nanowires were connected to superconducting NbTiN and a magnetic field was applied along the axis of nanowire. At the end of the nanowire, the charge carrier density was reduced appreciably making $\mu^2$ quite large. This made $(\mu^2 + \Delta^2)^{1/2}$ to be equal to Zeeman energy, i.e., $E_z$. The points in space which satisfy the condition $E_z = (\mu^2 + \Delta^2)^{1/2}$ host Majorana fermions as it represents the zero-energy bound states. These states can be detected by tunneling spectroscopy, in which they show zero energy peaks that are robust against change in the magnetic field. Another semiconducting material that qualifies the condition of high SOC is InAs. Rashaba spin-orbit coupling strength, $\alpha$, of both of the semiconductors is calculated in ref. [267] and [268]. This analysis is used to calculate the spin-orbit coupling energy by using the formula $E_{SO} = m^*\alpha^2/\hbar^2$ where $m^*$ is the effective mass of electrons and spin-orbit length is given by $\lambda_{SO} = \hbar^2/m^*\alpha$. Parameters like Lande g factor and electron effective mass are taken from ref. [269] and [270]. Thus, obtained parameters are listed in Table-5 which confirm the suitability of these semiconducting materials to obtain topological superconductivity. Aluminum (Al) is another superconductor that can be used in place of NbTiN. In recent studies topological superconductivity was also observed with its exotic phenomenon as an anomalous Josephson Effect in Al/InAs system [267,271,272].

## C2. Topological superconductivity at Superconductor and Topological Insulator Interface:

It was predicted in theoretical studies that Majorana Fermions can be found at the vortices of an interface between a simple s wave superconductor and a topological insulator [26]. As when an s-wave superconductor is deposited on the surface of a topological insulator, the equation for the excitation spectrum at the interface is given by

$$E_K = \pm[(\pm v|k| - \mu)^2 + \Delta_0^2]^{1/2} \qquad (2)$$

where $\mu$ is chemical potential and $\Delta_0$ is superconducting order parameter at absolute zero of s-wave superconductor. In special conditions when $\mu \gg \Delta_0$, this energy spectrum at low temperatures becomes very much similar to that for a $p_x + ip_y$ type superconductor or a spinless superconductor. It is well established that the vortex of a spinless superconductor hosts Majorana Fermions which can be probed by converting them into Dirac Fermions as Majorana Fermions are charge less particles [29,273-277]. Following this, superconductivity was observed in $Bi_2Se_3$ thin films deposited on $Si/SiO_2$ substrates [Fig. 22(a)]. The $Si/SiO_2$ served as the gate and contacts were made by the superconducting material of Al/Ti. A gate tuned supercurrent of 200 nA was observed at 30 mK at zero bias [278] as shown in Fig. 22(b). In another experiment that was made on $Bi_2Se_3$ nanoribbons deposited on a Si substrate, and contacts were made of superconducting tungsten [279], the same proximity induced superconductivity was observed at a relatively higher temperature close to the critical temperature of Tungsten. A critical current $I_c$ of 1.1 μA at 500 mK was observed in these measurements [279]. By following this feat, topological superconductivity was observed at the interface of heterostructures consisting of TI and a superconductor [253]. These heterostructures were grown by using Molecular Beam Epitaxy (MBE) deposition technique, taking $Bi_2Se_3$ as a TI and $NbSe_2$ as a superconductor. Also, proximity induced superconductivity was observed with other heterostructures as $Sn-Bi_2Se_3$ [280], $Bi_2Te_3-NbSe_2$ [281-283], $Bi_2Te_3-FeTe_{0.55}Se_{0.45}$ [284], $Sb_2Te_3-FeTe$ [285], $Pb-TlBiSe_2$ [286]. This method became a popular method to achieve topological superconductivity afterward. But the main problem with this method is that the interface between the topological insulator layer and the superconducting layer should be atomically sharp for the lattice mismatch between two layers of different materials. Interestingly, the use of 1-D structure in place of 2-D is found to be more useful to study topological superconductivity [287,288], as in 1-D geometry (nanowires) the surface states of a TI are fixed into a discrete set of modes and the number of modes remains odd while a parallel magnetic flux of h/2e threads the nanowire [289-291]. This makes sure that when the proximity effect induces superconductivity at the interface, the system would be a topological superconductor [287,292]. To achieve fully gapped superconductivity, the role of the device geometry is very important as the vortex winding across the nanowire is necessary to achieve fully gapped superconductor [293].

Apart from a topological insulator, there are several theoretical assumptions that topological superconductivity can be found at the interface of a topological semimetal (TSM) and a superconductor [294-297]. But on experimental grounds, there are very few reports which show topological superconductivity at the interface between a TSM and a superconductor. As per our knowledge topological superconductivity is observed at $Cd_3As_2$ and Nb interface [298] and also at Nb enriched surfaces of NbAs [299]. In TSMs, low energy bulk excitations and Fermi arc states exist simultaneously [300] which makes it difficult to determine how significant a role of the surface and bulk states play to achieve a superconducting state. Due to the existence of Fermi arc states, Majorana fermions appear as Majorana arcs [301] at the interface between TSM and superconductor.

Using of TIs nanowires instead of semiconducting nanowires have proved to be more useful to obtain topological superconductivity. This can be understood as, the necessity condition to achieve topological superconductivity is that the chemical potential should lie in the Zeeman gap opened in semiconducting nanowires due to application of magnetic field. In case of semiconducting nanowires, the magnitude of the Zeeman gap opened at the application of magnetic field of 1T is 1 meV [287], which is very low. While in the case of TIs nanowire's, chemical potential can lie anywhere inside the TI bulk gap which is 300 meV [287] for $Bi_2Se_3$ nanowire. Another advantage with TIs nanowire is that the induced superconducting gap is robust against any nonmagnetic impurity due to the present surface state in TI.

**C3. Topological superconductivity at interface of magnetic material and superconductor:**

Another system to realize topological superconductivity is an interface between the magnetic material and a superconductor. It was seen in Co nanowires on which contacts were made of superconducting tungsten, and topological superconductivity was observed at Co-W interface [260,261]. The proximity effect in ferromagnetic and superconducting hetero-structures is not similar to that is observed for normal metal and superconductor as there are two competing ordered states: one due to ferromagnetism and another due to superconductivity. The behavior of the Cooper pair wave function inside the ferromagnetic medium becomes a damped oscillator [288]. However, a long-range proximity effect is evident in the 1-D ferromagnetic junction from many experiments [302-304]. There are lots of difficulty in realizing topological superconductivity in these systems because of the low SOC presence at the interface, which suppresses the triplet pairing of Cooper pairs [261]. For this, it was suggested to use a layer of high SOC material between ferromagnet and superconductor. Also, this difficulty can be removed by using an antiferromagnetic insulator in place of ferromagnetic material [305]. This antiferromagnetic insulator provides Dirac states in its non-magnetic state which provides enough SOC that opens the gap at the interface or allows triplet pairing. This particular model of topological superconductivity can be achieved but this is still a hypothesis and yet to be experimentally realized. Recently it has been theoretically shown that superconductivity can be obtained at the surface of a topological insulator when it is coupled either of the ferromagnetic insulator or an antiferromagnetic insulator [306]. The reason for this superconductivity is the interfacial fermion magnon interaction. There are very few experimental reports that exist, on topological superconductivity at the interface of a superconductor and a magnetic material, so far it has been only shown theoretically. So, this field is open for new experiments that can be done to realize topological superconductivity. In comparison to the above three methods, the third one is the least popular, but it was used with using another layer of high SOC semiconducting material between the superconductor and magnetic material as in ref. [27].

**Conclusion:**

Currently, search for materials that can show topological superconductivity is one the most imperative topics of interest due to their importance in quantum computation devices. Majorana fermions have a direct relation with topological superconductivity as established through numerous signatures of their presence. It shows the importance of topological superconductors as these can host these particles in their bulk form. These fermions are a key part of fault-tolerant quantum computing. This review mainly describes the experimental progress on materials that are supposed to show topological superconductivity and confirms that the bulk materials stand out to be a better platform for topological superconductivity. There are some unanswered questions in regards to topological superconductivity in doped-bulk materials as there is a dilemma on how intercalation of some specific elements induces superconductivity in TIs. The topological superconductivity in doped topological insulators as well as in some bulk superconductors is a more convenient way to realize topological superconducting states. Other methods to attain topological superconductivity such as the application of pressure and heterostructures were also discussed. For pressure-induced superconductivity in topological materials, it is hard to say whether the material remains in a topological non-trivial state after application of such high pressure as the application of pressure completely changes the phase and atomic structures of material. For heterostructures, the principal challenge is the need of interface between the two materials to be very clean and there should be very little lattice mismatch- this is very tedious work. The results of μSR studies made on bulk topological superconductors are also discussed to determine the role of pairing symmetry in superconductivity and how the pairing symmetry in $Sr_xBi_2Se_3$ is different from that in $Nb_xBi_2Se_3$ and $Cu_xBi_2Se_3$. This review aims to help readers to obtain information about recent experimental advances in the field of topological superconductivity to motivate them to make advancements in this exciting field.

**Acknowledgement:**

The authors would like to thank Director NPL for his keen interest and encouragement. M.M. Sharma and N. K. Karn would like to thank CSIR for the research fellowship, and Prince Sharma would like to thank UGC for fellowship support. M.M. Sharma, Prince Sharma and N. K. Karn are also thankful to AcSIR for Ph.D. registration.

**Table-1**

**Superconductivity parameters of doped topological Insulators:**

| Topological Insulator | Stoichiometry x | $T_c$ | Volume Fraction |
|---|---|---|---|
| $Cu_xBi_2Se_3$ [104,112,115] | 0.12 < x < 0.60 | 3.8K | 20% - 50% |
| $Sr_xBi_2Se_3$ [107] | 0.06 < x < 0.1 | 2.8K | Upto 100% |
| $Nb_xBi_2Se_3$ [105,108] | x = 0.25 | 2.5-3.4K | 60% - 95% |
| $PdBi_2Te_3$ [110,111,148] | 0.12< x < 1 | 5.5K | 1% |
| $Tl_xBi_2Te_3$ [150] | x = 0.6 | 2.2K | 95% |

**Table-2**

**Effect of pressure on phase and $T_c$ of Topological Insulators:**

| Topological Insulator | Pressure | Crystal Structure | Critical Temperature ($T_c$) |
|---|---|---|---|
| $Bi_2Se_3$ [229,230] | Ambient | R -3 m | Non-Superconducting |
| | 12GPa | C 2/m | 4.4K |
| | 20GPa | C 2/c | 8.2K |
| | 29GPa | I m -3 m | 6.5K |
| $Bi_2Te_3$ [231] | Ambient | R -3 m | Non-superconducting |
| | 4GPA | R -3 m | 3 K |
| | 8GPa | C 2/m | 8 K |
| $Sb_2Te_3$ [232,233] | Ambient | R -3 m | Non-superconducting |
| | 4GPa | R -3 m | 3 K |
| | 9.3GPa | C 2/m | 7 K |
| | 19.6GPa | I m -3 m | 7 K |

## Table-3

**Effect of pressure on phase and $T_c$ of Weyl Semi Metal:**

| Weyl Semimetal | Pressure | Crystal Structure | Critical Temperature ($T_c$) |
|---|---|---|---|
| MoTe$_2$ [213] | Ambient | P2$_1$/m | 0.10K |
| | 1GPa | - | 5K |
| | 11.7GPa | - | 8.2K |
| WTe$_2$ [234] | Ambient | P m n 2$_1$ | Non-Superconducting |
| | 2.5GPa | - | 3.1K |
| | 16.1GPa | - | 7K |
| TaP [235] | Ambient | I4$_1$ md | Non-Superconducting |
| | 70GPa | P -6 m 2 | 3.1K |
| | 100GPa | - | 1.7K |

## Table-4

**Effect of pressure on phase and $T_c$ of Dirac Semi Metal:**

| Dirac Semimetal | Pressure | Phase | Critical Temperature ($T_c$) |
|---|---|---|---|
| Cd$_3$As$_2$ [236] | Ambient | I4$_1$/acd | Non-Superconducting |
| | 8.5GPa | I4$_1$/acd | 2K |
| | 21.3GPa | P2$_1$/c | 4.1K |
| ZrTe$_5$ [237] | Cmcm | C m c m | Non-Superconducting |
| | 6.2GPa | C2/m | 1.8K |
| | 21.2GPa | P-1 | 6K |
| HfTe$_5$ [238] | Ambient | C m c m | Non-Superconducting |
| | 5GPa | C2/m | 1.8K |
| | 20GPa | P-1 | 4.8K |

## Table-5

**List of candidates of topological superconductivity with their pairing symmetry and their category based on Time-Reversal Symmetry obtained from available μSR data.**

| Nature of Time-Reversal Symmetry | Material | Pairing symmetry | Type of Topological material and Superconductivity |
|---|---|---|---|
| Time Reversal Symmetry Broken | $Sr_2RuO_4$ | Spin Singlet [53] | Unconventional Superconductor |
| | $Sr_xBi_2Se_3$ | Spin Triplet [244,245] | Doped Topological Insulator |
| Time Reversal Symmetry Preserved | BiPd | Mixed pairing Symmetry [67,68] | Non-centrosymmetric Superconductor |
| | $PbTaX_2$ (X=Se,S) | Spin singlet [242] | Nodal line semimetal and non-centrosymmetric superconductor |
| | NbC & TaC | Spin singlet [94] | Dirac type-II semimetal and centrosymmetric superconductor |
| | $M_xBi_2Se_3$ (M= Cu,Nb) | Spin triplet [140,243] | Doped Topological Insulator |
| | $In_xSn_{1-x}Te$ | Singlet Pairing [158] | Doped Crystalline Insulator |
| | $MoTe_2$ | Singlet Pairing [246] | Dirac Type-II Semi-metal |
| | $PdTe_2$ | Singlet Pairing [247,248] | Dirac Type-II Semi-metal |

## Table-6

**Parameters related to Spin-Orbit Coupling of InAs and InSb semiconductors:**

| Properties | Semiconductor | |
|---|---|---|
| | InAs | InSb |
| Lande g factor (g) | 8-15 [269,271] | 40-50 [251,260] |
| Effective mass of electron ($m^*$) | 0.023 $m_e$ [272] | 0.015 $m_e$ [262] |
| Spin-Orbit Coupling Strength ($\alpha$) | 0.2-0.8 eV Å [253,269] | 0.2-1 eV Å [251,270] |
| Spin-Orbit Coupling Energy ($E_{SO}$) | 0.05-1 meV [253,269] | 0.25-1 meV [251,270] |
| Spin-Orbit Coupling Length ($\lambda$) | 180-40 nm.[ 253,269] | 230-50 nm [251,270] |

**Figures Caption:**

Figure 1. An illustration of how Majorana particle excitations emerge from the interaction of electrons and holes with cooper pairs in p wave superconductors making electrons and holes indistinguishable. Adapted from ref. [28].

Figure 2. Pictorial view of quantized thermal hall effect in a p wave superconductor showing heat current carried by Majorana edge modes. Adapted from ref. [39].

Figure 3. STM image of vortex core pinned at the step edge in BiPd under a magnetic field of 53mT and zero energy peak in STS measurements at the center of the core. Reprinted from ref. [68].

Figure 4. In-plane and out of plane magnetoresistance of $PbTaSe_2$ at 5K up to 2T. Reprinted from Ref. [77]; copyright (2016) by the American Physical Society.

Figure 5(a). Rietveld refined XRD pattern of NbC in which, right-hand side inset is showing the FC and ZFC measurements at 8Oe and left-hand side inset is showing variation of $T_c$ with respect to lattice constant. (b). DFT calculations based bulk electronic band structure of NbC. (c) Evolution of Wannier charge centers in various K-planes. Adapted from ref. [97] with permission of Springer nature.

Figure 6. Specific heat vs Temperature plot of $Cu_xBi_2Se_3$. Reprinted from Ref. [112]; copyright (2011) by the American Physical Society.

Figure 7. (a) ARPES image of $Sr_xBi_2Se_3$ for x=0 and x=0.05 showing shift in Dirac point. Reprinted from ref. [128], with the permission of AIP Publishing. (b) ARPES image of $Cu_xBi_2Se_3$ for x=0 and x= 0.10 showing shift in Dirac point. Reprinted from Ref. [129]; copyright (2012) by the American Physical Society.

Figure 8. DC magnetization plots under FC and ZFC conditions of $Nb_{0.25}Bi_2Se_3$ in presence of a magnetic field of 20 Oe inset show M-H curve down to 2K. Reprinted from ref. [108], with permission of Springer nature.

Figure 9. (a) Optimized heat treatment profile of $Nb_{0.25}Bi_2Se_3$. Adapted from ref. [141]. (b) XRD pattern of $Nb_{0.05}Bi_2Se_3$ from 10 to 50° showing presence of two phases red one in of Bi2Se3 and black one is of NbBiSe3, inset shows spreaded plot around 14° to 15° for x= 0, 0.05, 0.2, 0.3, 0.6, 0.7. Reprinted from Ref. [106]; copyright (2017) by the American Physical Society. (c) XRD pattern of Nb0.25Bi2Se3 from 0 to 50° showing presence of single phase of $Bi_2Se_3$. Adapted from ref. [141], with permission of Springer nature. (d) Raman Spectra Comparison of $Bi_2Se_3$ and $Nb_{0.25}Bi_2Se_3$, Reprinted from ref. [108], with permission of Springer nature.

Figure 10. Pressure dependence of Tc of $Sr_xBi_2Se_3$, $Nb_xBi_2Se_3$ and $Cu_xBi_2Se_3$. Reprinted from ref. [143] with permission from Elsevier.

Figure 11. Quantum Oscillations (dHvA effect) with multiple frequencies observed in $Nb_xBi_2Se_3$. Reprinted from Ref. [144]; copyright (2016) by the American Physical Society.

Figure 12. Pressure dependent R-T plots of $Sn_{0.5}In_{0.5}Te$. Republished with permission of IOP Publishing, from ref. [161].

Figure 13. Rietveld refined XRD pattern of $Sn_{1-}Sb_x$ (x=0.4,0.5,0.6) and their respective unit cells are shown on the right-hand side of XRD patterns. Reprinted from ref. [182] with permission from Elsevier.

Figure 14. Energy band diagram of (a) $Sn_{0.4}Sb_{0.6}$ (b) $Sn_{0.5}Sb_{0.5}$ (c) $Sn_{0.6}Sb_{0.4}$, based on first-principle calculations without SOC and with SOC. Reprinted from ref. [182] with permission from Elsevier.

Figure 15. FC-ZFC plots of $Sn_xSb_{1-x}$ (x=0.4, 0.5, 0.6) in presence of 10 Oe magnetic field Reprinted from ref. [182] with permission from Elsevier. inset is showing isothermal M-H plots up to 4K for $Sn_{0.4}Sb_{0.6}$. Reprinted from ref. [178] with permission from Elsevier.

Figure 16(a) Rieltveld refined XRD pattern of SnAs in which, inset is showing the unit cell of SnAs. (b) FC and ZFC plot of SnAs under a magnetic field of 12 Oe, inset is showing M-H plot at 2K of the same. Reprinted from ref. [184] with permission from Elsevier.

Figure 17(a). DFT calculated electronic band structure along with Density of states of SnAs with and without SOC. (b) Surface states in SnAs mapped at the Γ point. (c) Fermi surface of SnAs. (d) Evolution of Wannier charge centers in various K-planes in SnAs. Reprinted from ref. [184] with permission from Elsevier.

Figure 18. (a) STM image of vortex core of FeSe$_{0.45}$Te$_{0.55}$. Reprinted from [202], reprinted with permission from AAAS. (b) Zero energy peaks observed in STS measurements for different distances from the center of the vortex core. Reprinted from [202], reprinted with permission from AAAS.

Figure 19. Plots of variation in relaxation rates viz. σ$_{ZF}$ and λ$_{ZF}$ with respect to temperature for Sr$_x$Bi$_2$Se$_3$. Reprinted from Ref. [244]; copyright (2019) by the American Physical Society

Figure 20. TF-μSR plots of Sr$_x$Bi$_2$Se$_3$ at (a) 0.09K & (b) 3.6K (c) Depolarization vs Temperature plot of Sr$_x$Bi$_2$Se$_3$ and (d) Internal Field vs Temperature plot of Sr$_x$Bi$_2$Se$_3$. Reprinted from Ref. [244]; copyright (2019) by the American Physical Society.

Figure 21. (a) Schematic of heterostructures proposed for Topological Superconductor. Reprinted from Ref. [27]; copyright (2010) by the American Physical Society. (b) E-K diagram with Spin-Orbit interaction in the absence of magnetic field and presence of the magnetic field. Adapted from ref. [266].

Figure 22. (a) Schematic of the complete device to observe the gate tuned supercurrent at TI/superconductor junction deposited on Si/SiO$_2$ substrate. Adapted from ref. [278] (b) Supercurrent observed at TI/Superconductor junction at a temperature of 30 mK and Zero Bias. Reprinted from ref. [278].

Fig. 1

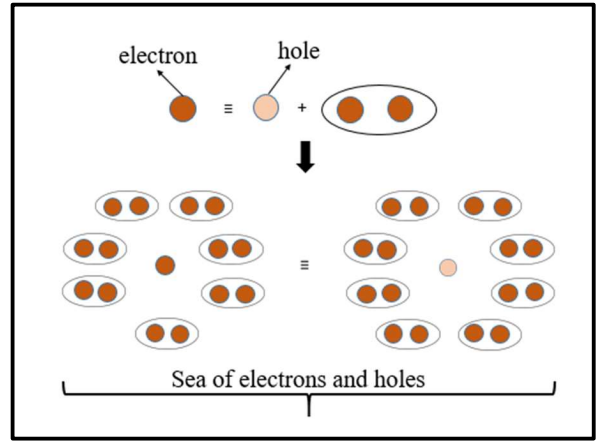

Fig. 2

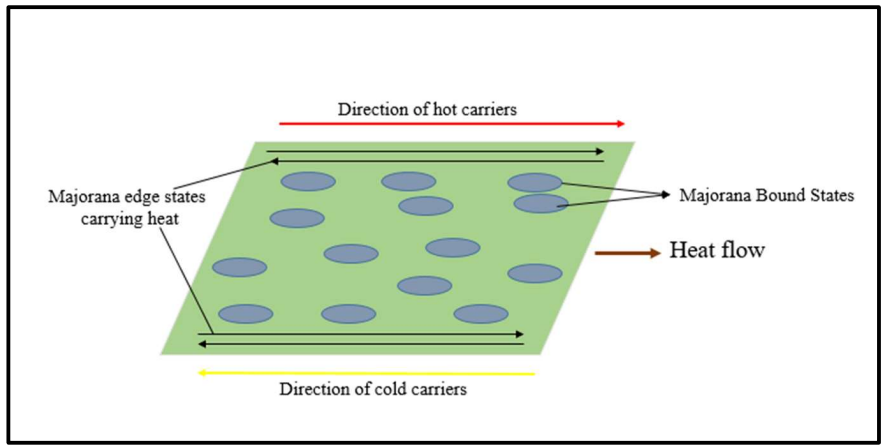

Fig. 3

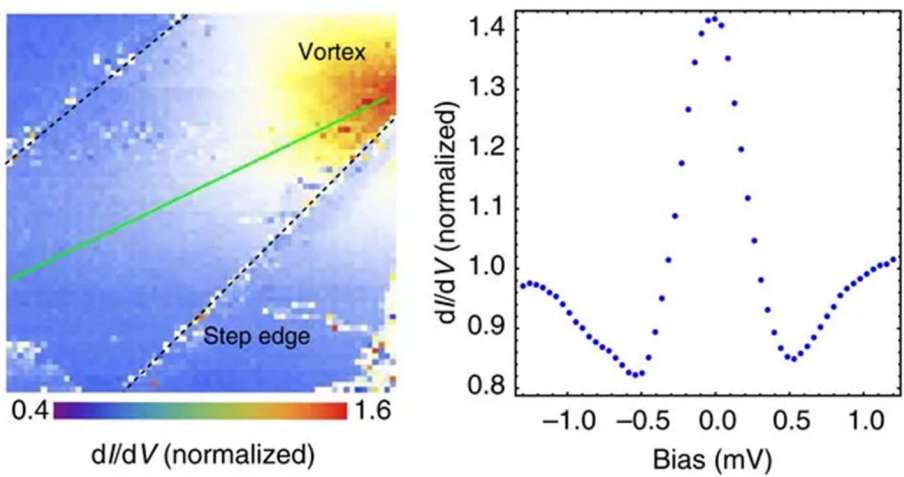

Fig. 4

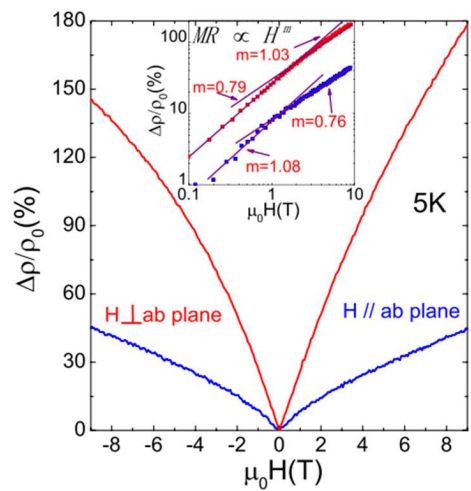

Fig. 5(a)

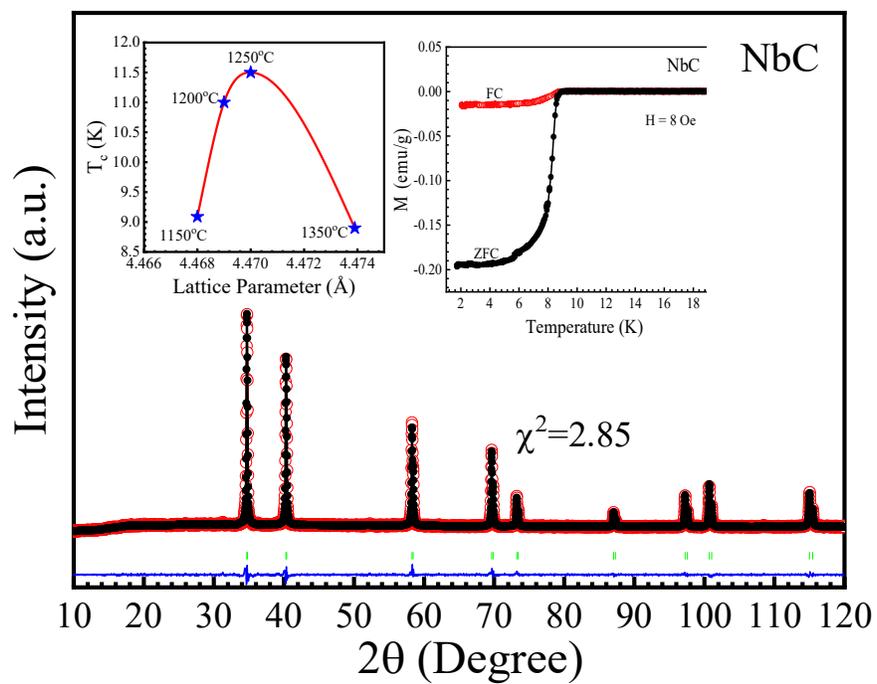

Fig. 5 (b)

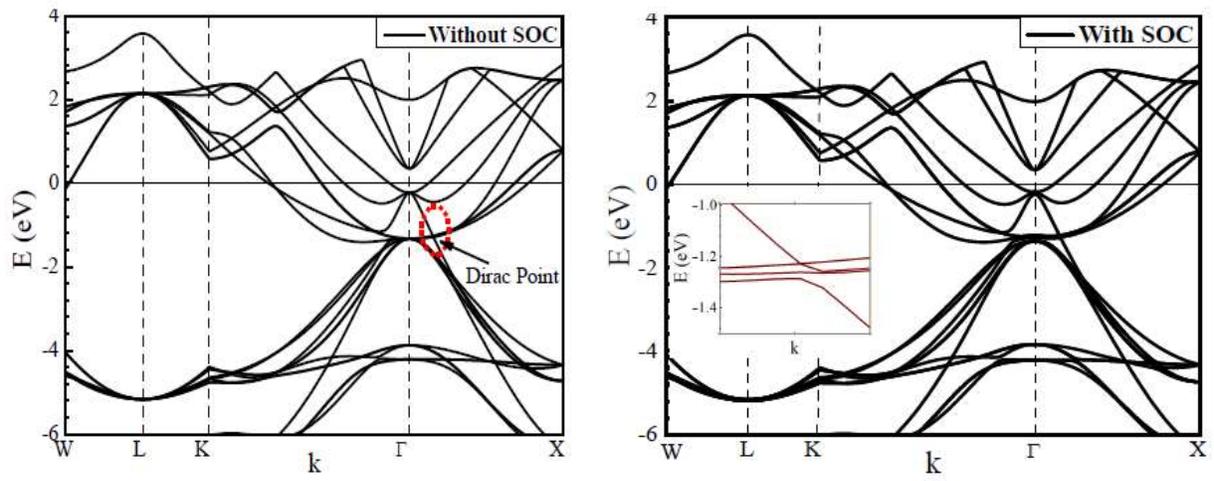

Fig. 5(c)

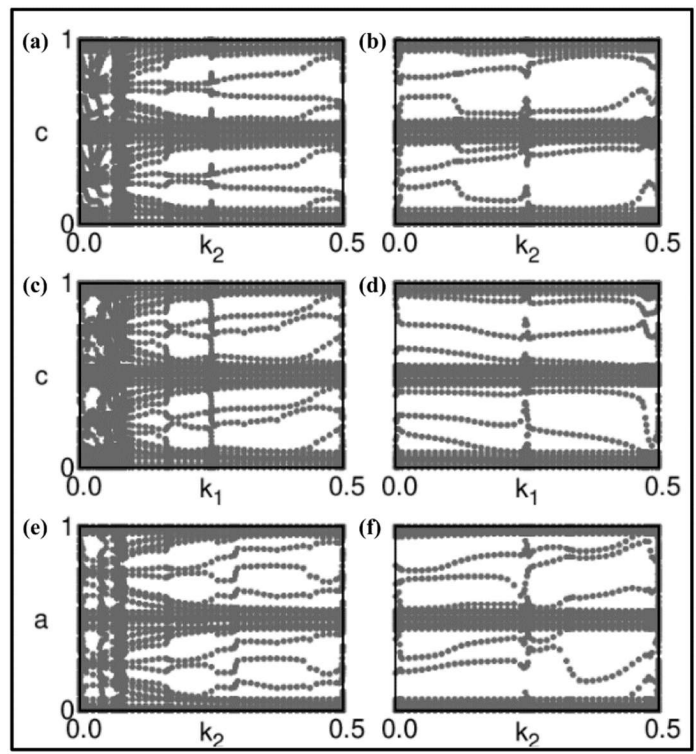

Fig. 6

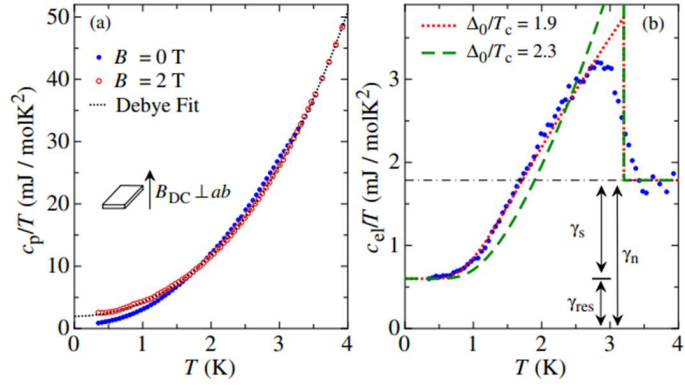

Fig. 7(a)

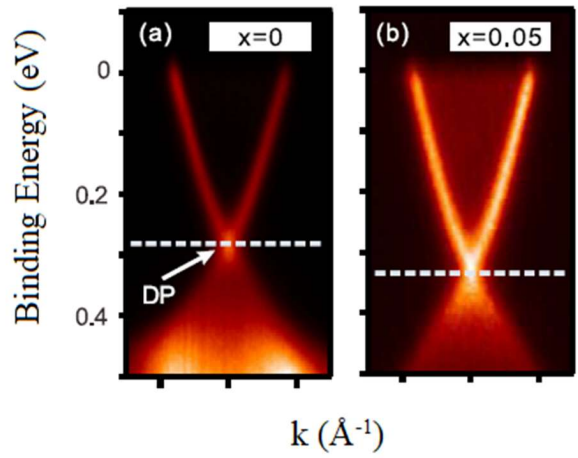

Fig. 7(b)

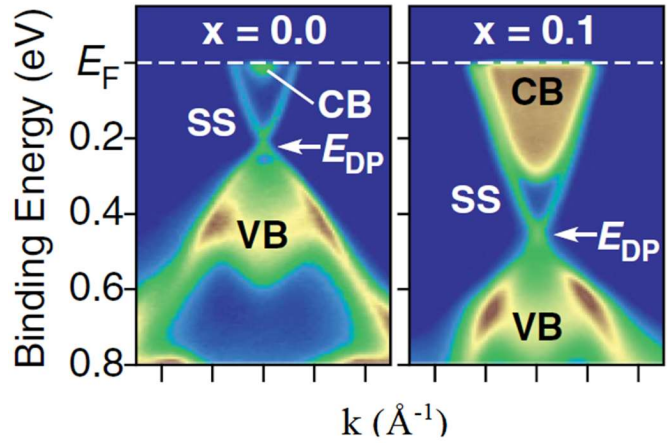

Fig. 8

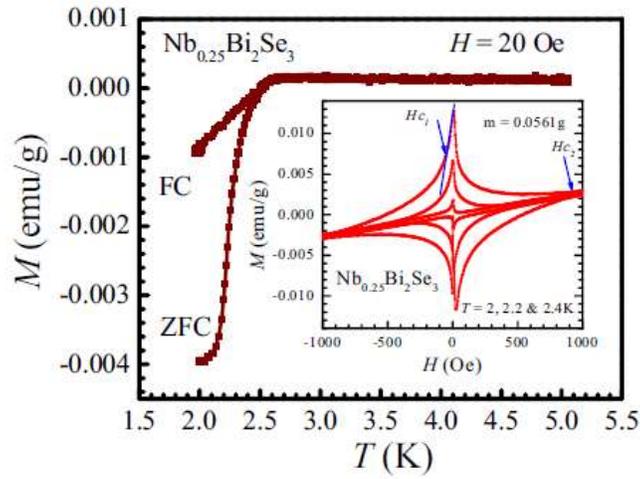

Fig. 9(a)

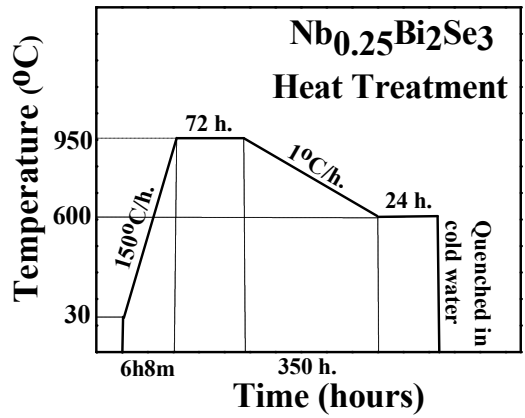

Fig. 9(b)

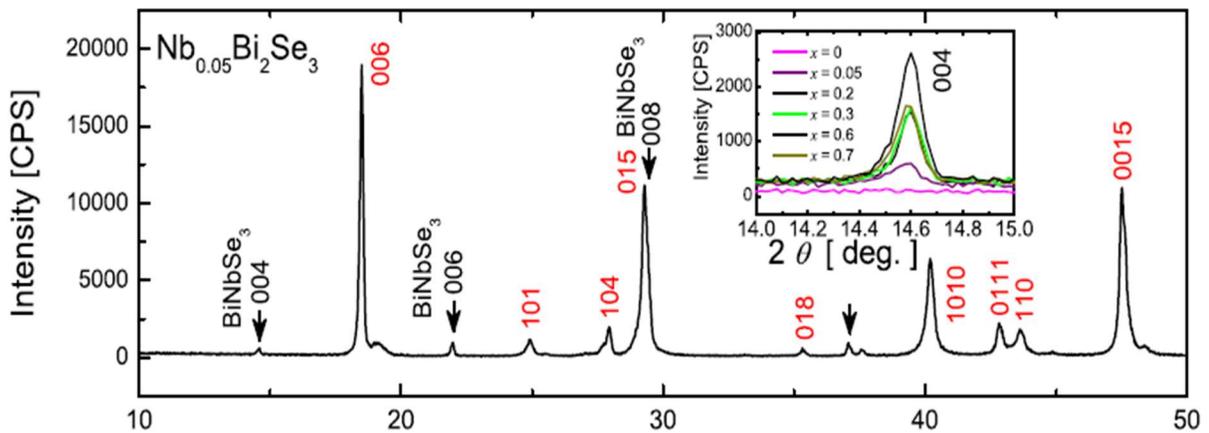

Fig. 9(c)

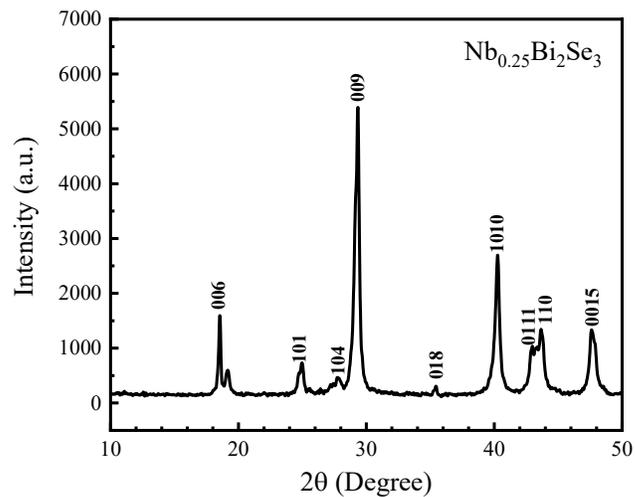

Fig. 9(d)

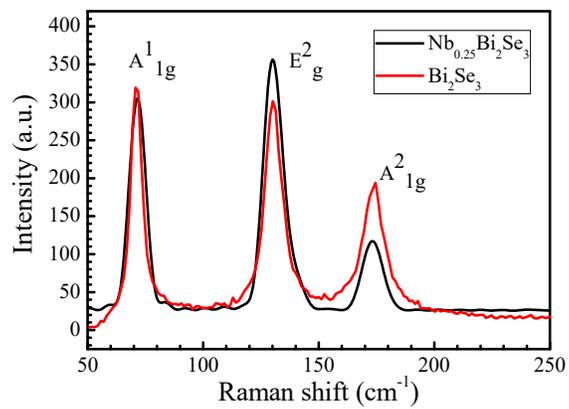

Fig. 10

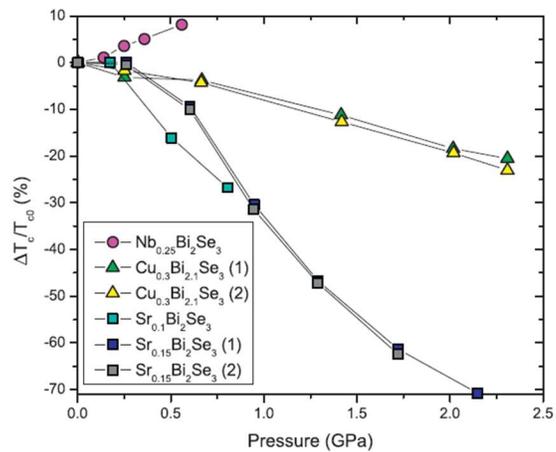

Fig. 11

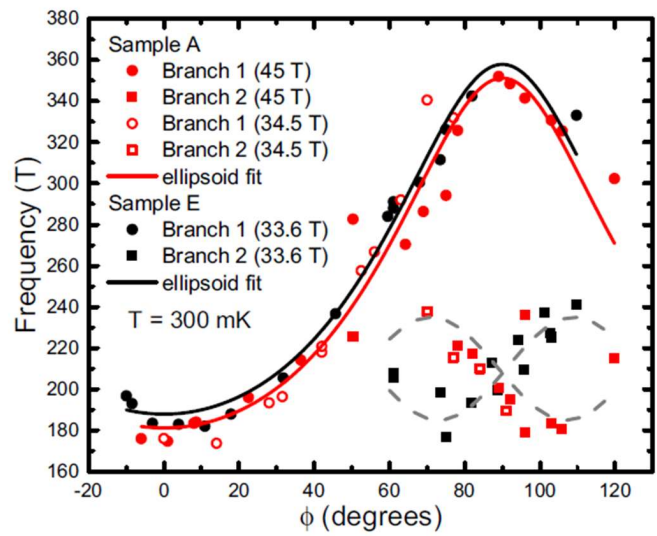

Fig. 12

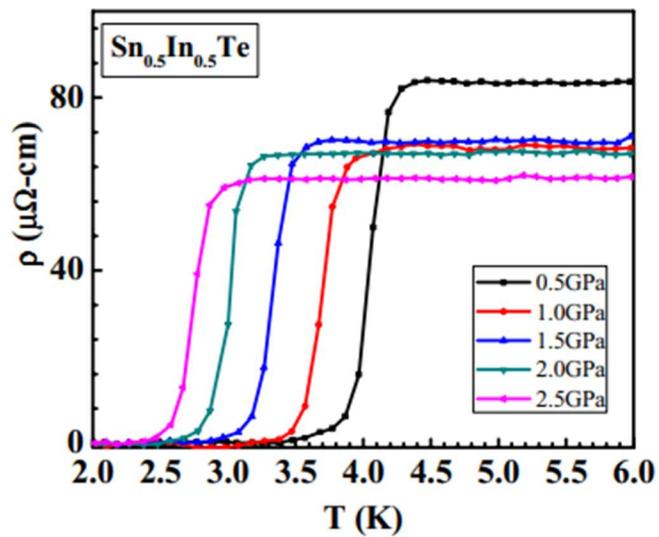

Fig. 13

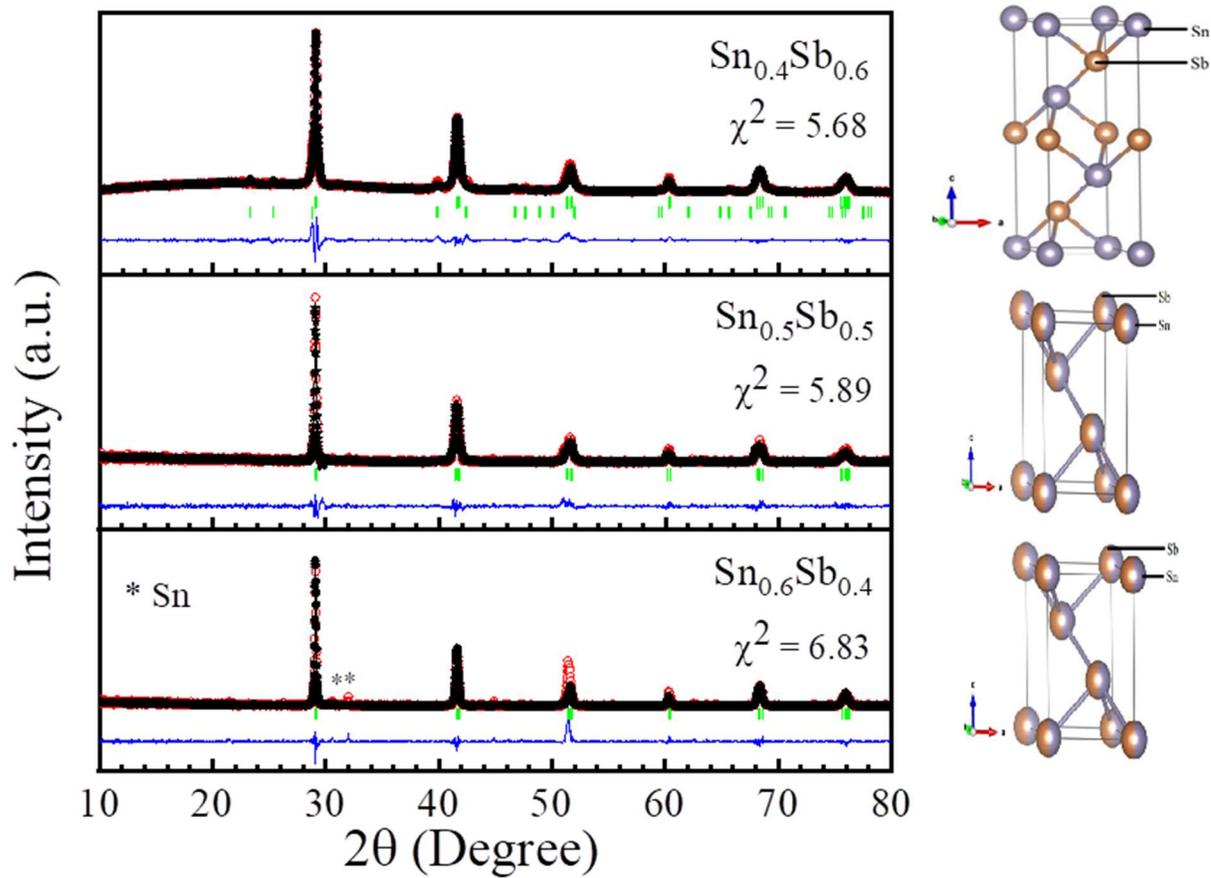

Fig. 14

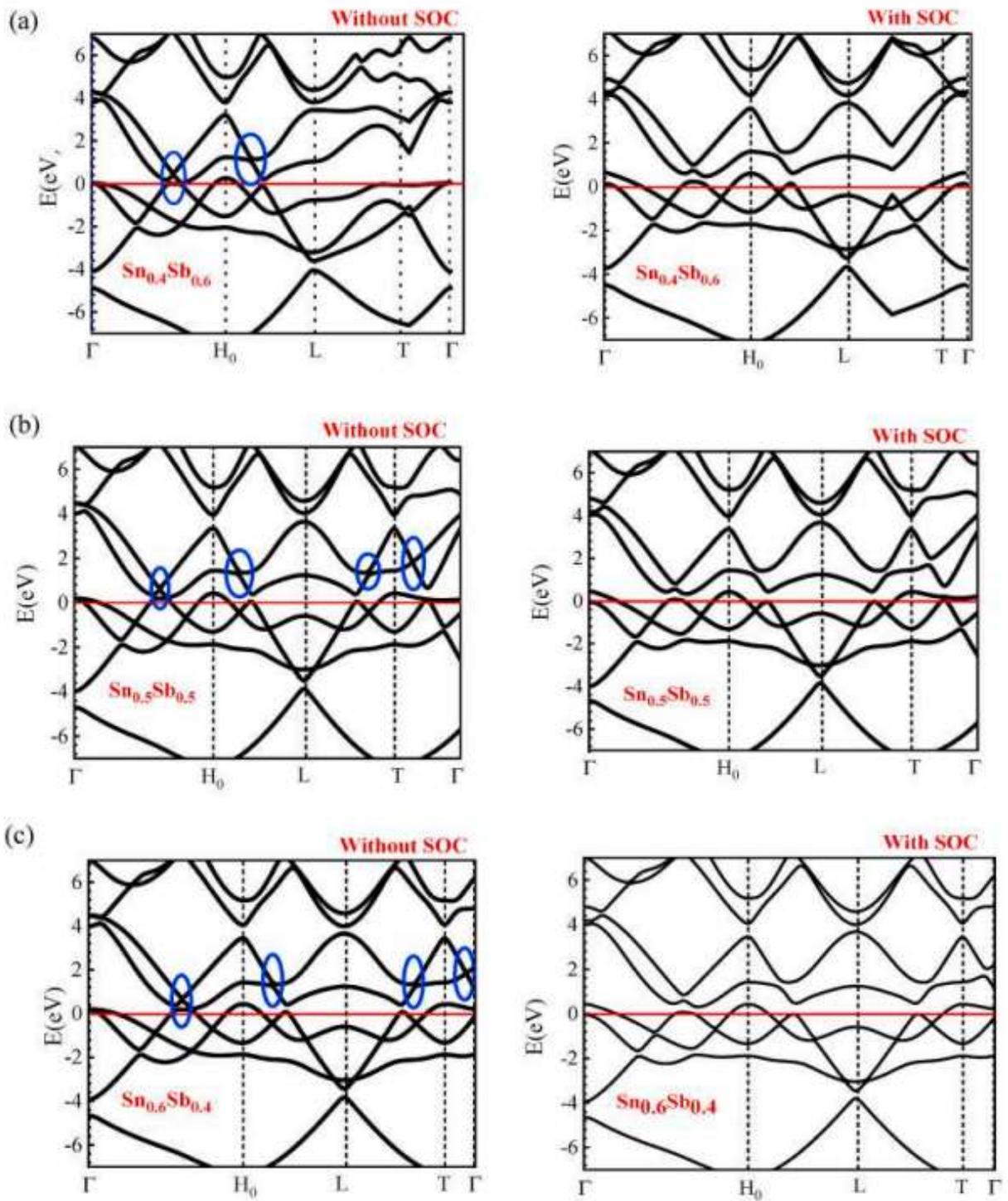

Fig. 15

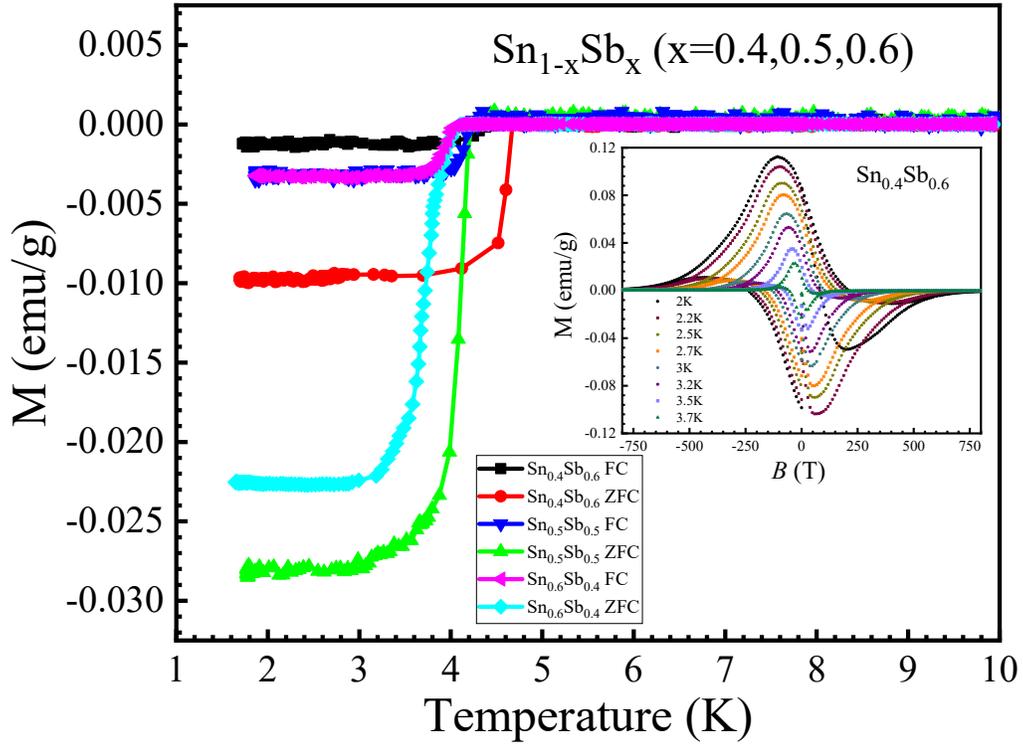

Fig. 16(a)

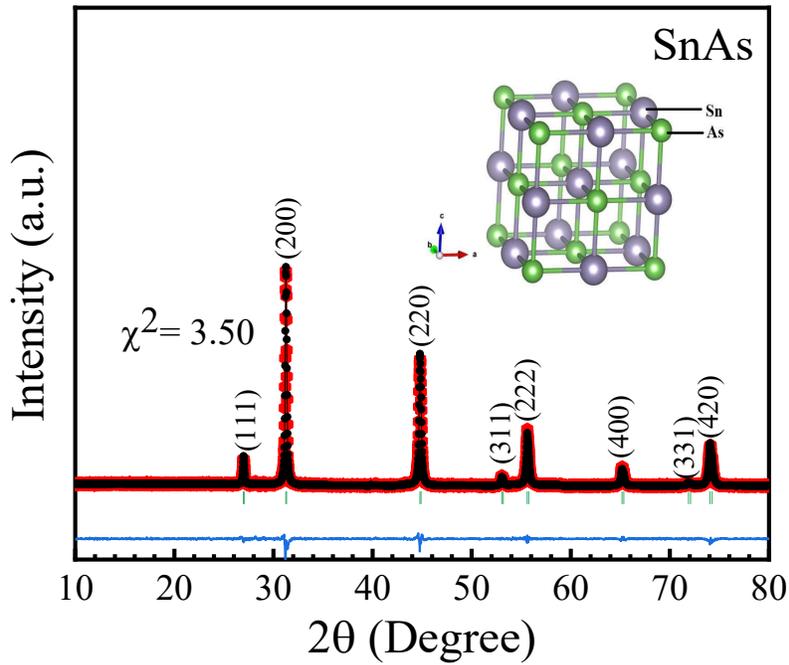

Fig. 16(b)

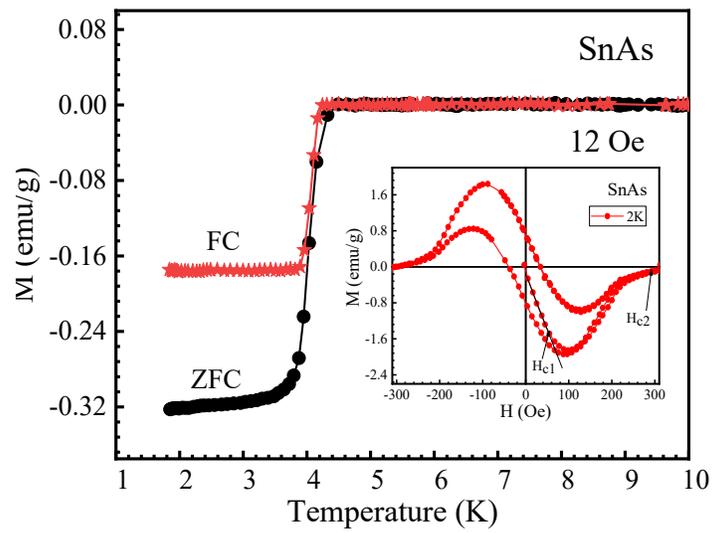

Fig. 17(a)

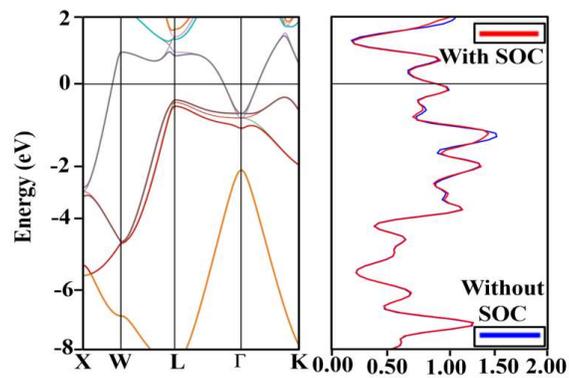

Fig. 17(b)

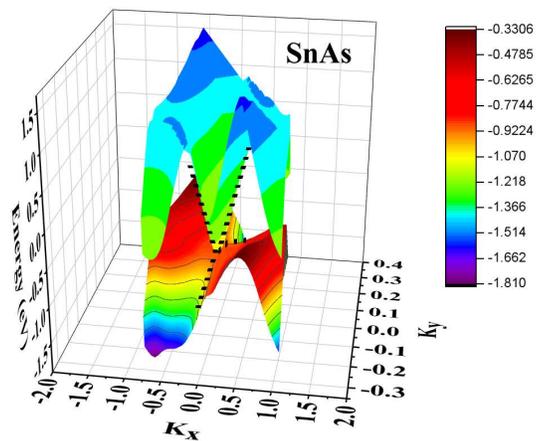

Fig. 17(c)

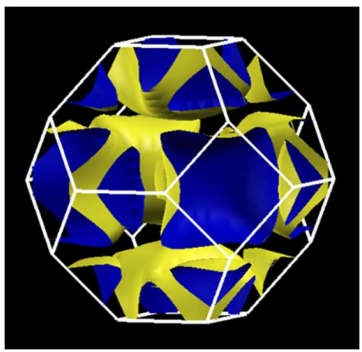

Fig. 17(d)

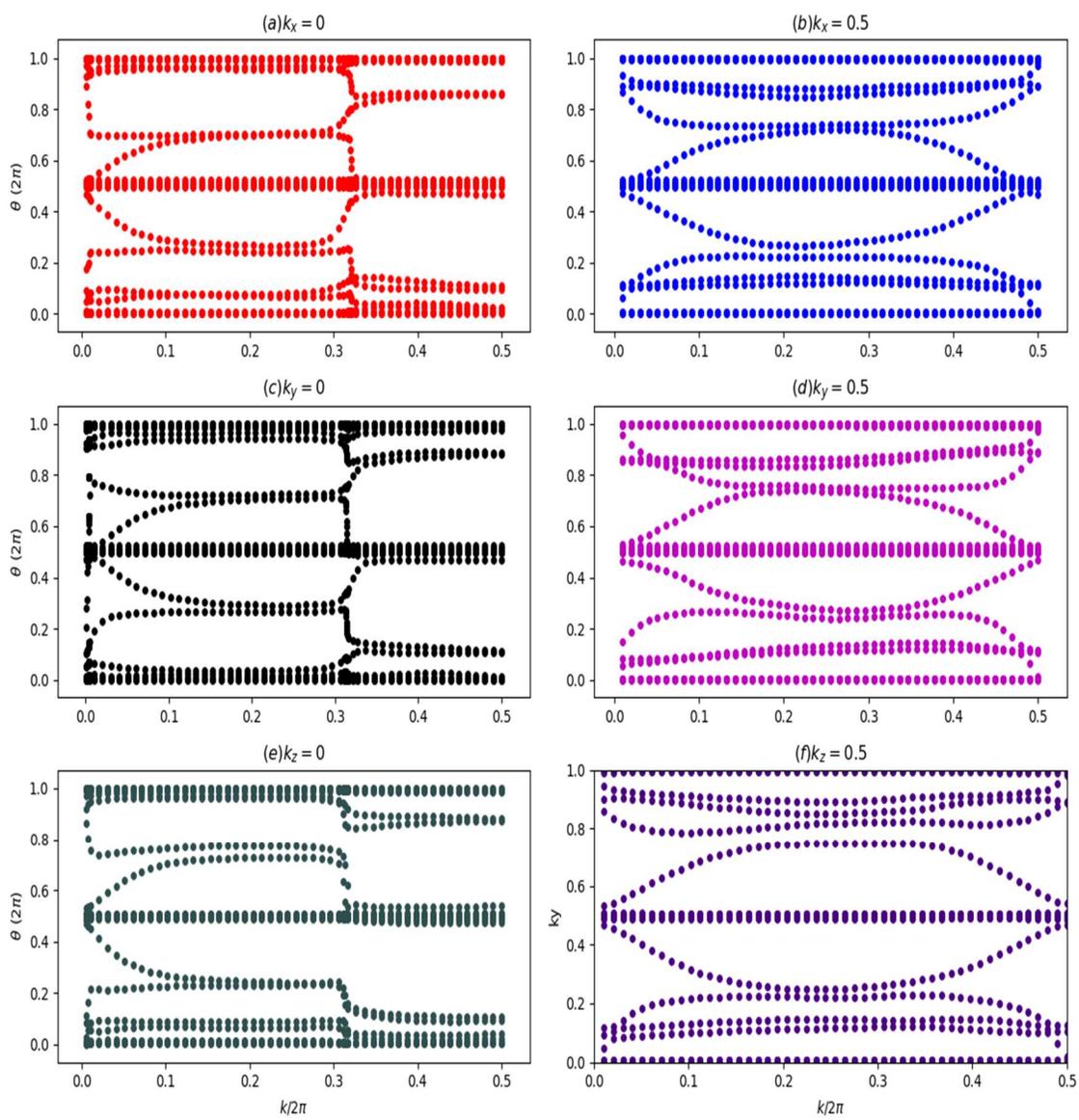

Fig. 18

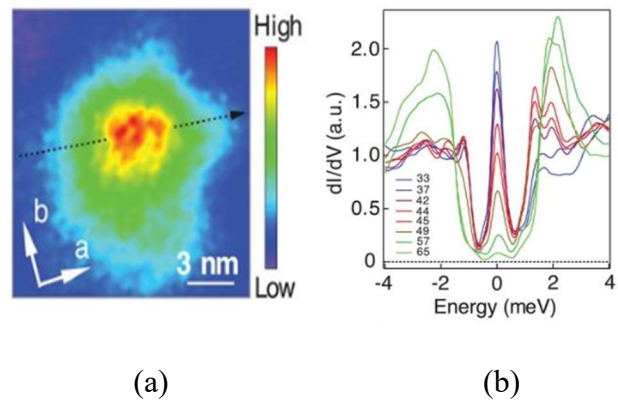

(a)                  (b)

Fig. 19

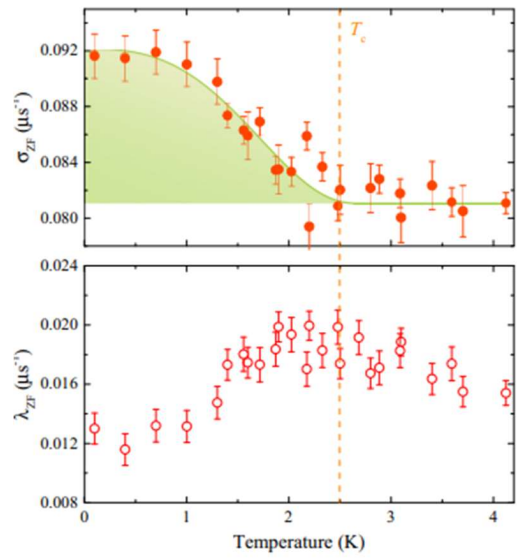

Fig. 20

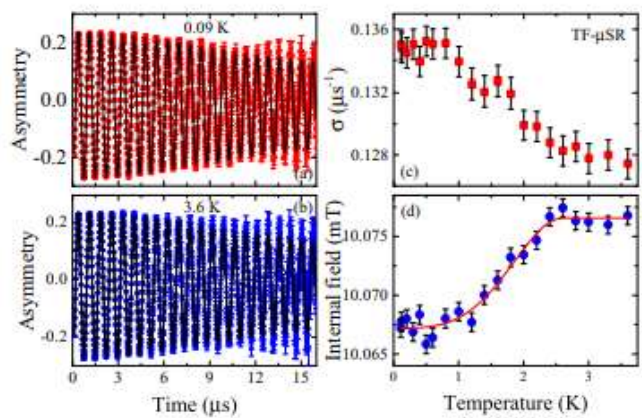

Fig. 21(a)

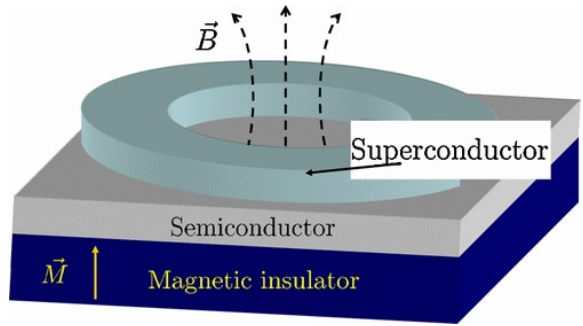

Fig. 21(b)

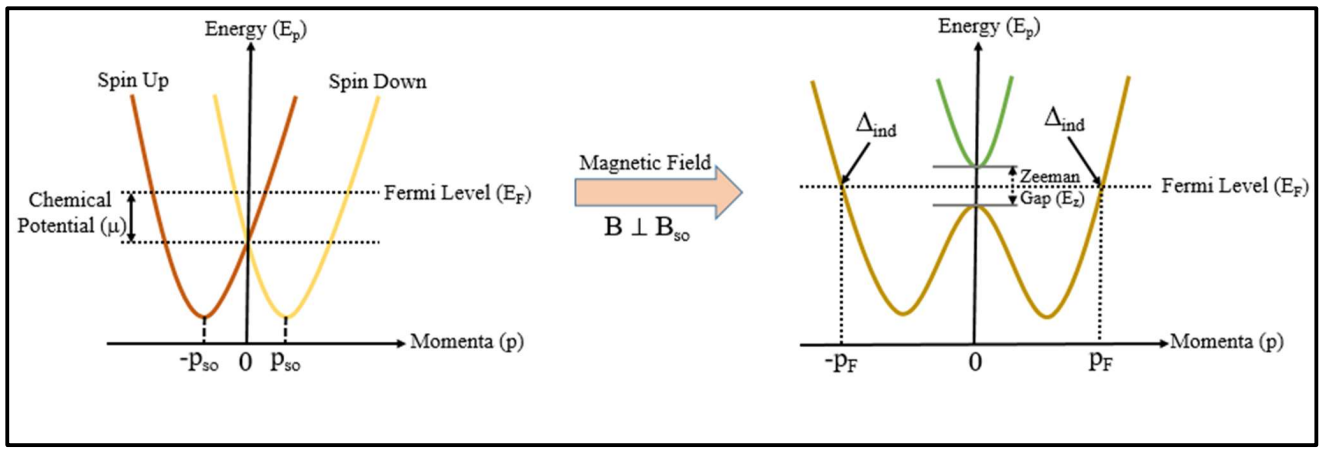

Fig. 22(a)

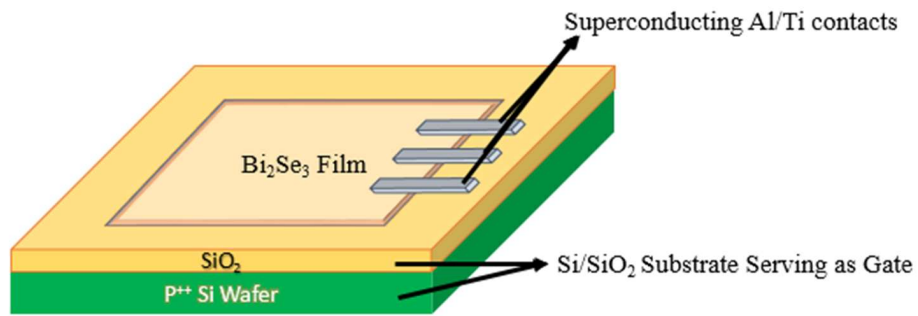

Fig. 22(b)

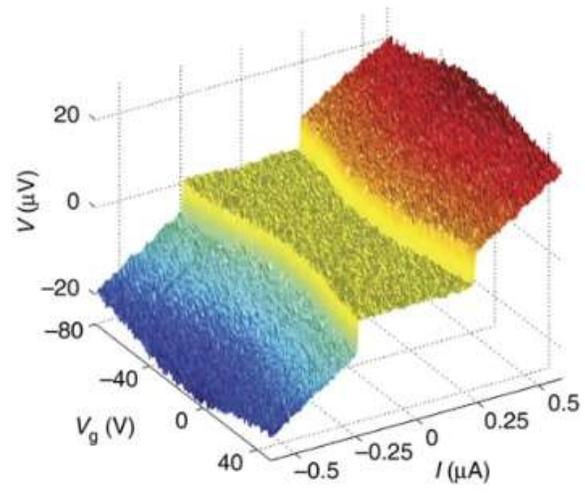